\documentclass[preprint,12pt]{elsarticle2}
\usepackage[margin=25truemm]{geometry}

\usepackage{amsmath,amssymb}
\usepackage{amsthm}

\theoremstyle{plain}
\newtheorem{thm}{Theorem}
\newtheorem{lem}{Lemma}

\usepackage{bm}

\usepackage{subfigure}
\usepackage{comment}
\usepackage{subfiles}
\usepackage{color}

\usepackage{url}

\usepackage{lineno}
\newcommand*\patchAmsMathEnvironmentForLineno[1]{%
  \expandafter\let\csname old#1\expandafter\endcsname\csname #1\endcsname
  \expandafter\let\csname oldend#1\expandafter\endcsname\csname end#1\endcsname
  \renewenvironment{#1}%
     {\linenomath\csname old#1\endcsname}%
     {\csname oldend#1\endcsname\endlinenomath}}%
\newcommand*\patchBothAmsMathEnvironmentsForLineno[1]{%
  \patchAmsMathEnvironmentForLineno{#1}%
  \patchAmsMathEnvironmentForLineno{#1*}}%
\AtBeginDocument{%
\patchBothAmsMathEnvironmentsForLineno{equation}%
\patchBothAmsMathEnvironmentsForLineno{align}%
\patchBothAmsMathEnvironmentsForLineno{flalign}%
\patchBothAmsMathEnvironmentsForLineno{alignat}%
\patchBothAmsMathEnvironmentsForLineno{gather}%
\patchBothAmsMathEnvironmentsForLineno{multline}%
}

\journal{an international journal}

\begin{document}

\begin{frontmatter}



\title{Modeling pedestrian fundamental diagram based on\\ Directional Statistics}

\author[label1]{Kota Nagasaki}
\affiliation[label1]{organization={Department of Civil and Environmental Engineering, School of Environment and Society, Tokyo Institute of Technology},
            addressline={2-12-1 Ookayama, Meguro-city},
            city={Tokyo},
            postcode={152-8550},
            country={Japan}}
\author[label1]{Keiichiro Fujiya}
\author[label1]{Toru Seo\corref{cor1}}
\cortext[cor1]{The corresponding author. Email: \url{seo.t.aa@m.titech.ac.jp}}


\begin{abstract}
	Understanding pedestrian dynamics is crucial for appropriately designing pedestrian spaces. 
	The pedestrian fundamental diagram (FD), which describes the relationship between pedestrian flow and density within a given space, characterizes these dynamics.
	Pedestrian FDs are significantly influenced by the flow type, such as uni-directional, bi-directional, and crossing flows.
	However, to the authors' knowledge, generalized pedestrian FDs that are applicable to various flow types have not been proposed.
	This may be due to the difficulty of using statistical methods to characterize the flow types.
	The flow types significantly depend on the angles of pedestrian movement; however, these angles cannot be processed by standard statistics due to their periodicity.
	In this study, we propose a comprehensive model for pedestrian FDs that can describe the pedestrian dynamics for various flow types by applying Directional Statistics.
	First, we develop a novel statistic describing the pedestrian flow type solely from pedestrian trajectory data using Directional Statistics.
	Then, we formulate a comprehensive pedestrian FD model that can be applied to various flow types by incorporating the proposed statistics into a traditional pedestrian FD model.
	The proposed model was validated using actual pedestrian trajectory data.
	The results confirmed that the model effectively represents the essential nature of pedestrian dynamics, such as the capacity reduction due to conflict of crossing flows and the capacity improvement due to the lane formation in bi-directional flows.
\end{abstract}




\end{frontmatter}


\section{Introduction}\label{chap:intro}

Understanding pedestrian dynamics is essential for appropriately designing pedestrian spaces, such as corridors and public squares.
Flow, density, and speed are the most basic state variables of pedestrian flow.
A fundamental diagram (FD), which is the relationships between these variables, characterizes pedestrian flow and determines the important features such as the flow capacity.
In addition, FDs are useful for evaluating pedestrian flow models and developing dynamic simulation models \cite{vanumu2017fundamental}.
For these reasons, FDs are important and form the basis for the design of pedestrian spaces.

The specific natures of pedestrian flow should be taken into account in the modeling of the pedestrian FD, since FD was initially developed for vehicular flow.
May \cite{may1990traffic} showed similarities and differences between pedestrian and vehicular flows.
One of the most essential differences is that pedestrian flow is not unidirectional.
Thus, pedestrians often conflict with others' movement, and their collective behavior becomes very complicated.

This multi-directionality of pedestrian flow gives rise to complicated phenomena.
For example, if two pedestrian flows intersect at a {\it 90-degree angles}, capacity and speed will decrease due to potential collisions.
However, if two pedestrian flows pass each other at a {\it 180-degrees angle}, capacity and speed will remain the same or even increase due to the self-organization phenomena called {\it lane formation} \cite{feliciani2016empirical}.
If pedestrian flows with various angles intersects, it is not easy to predict the resulting flow.
Vanume et al.~\cite{vanumu2017fundamental} reviewed that this kind of {\it flow type} of pedestrian affects pedestrian FDs in a complicated manner.

\begin{figure}[tbp]
    \centering
    \includegraphics[width=15cm]{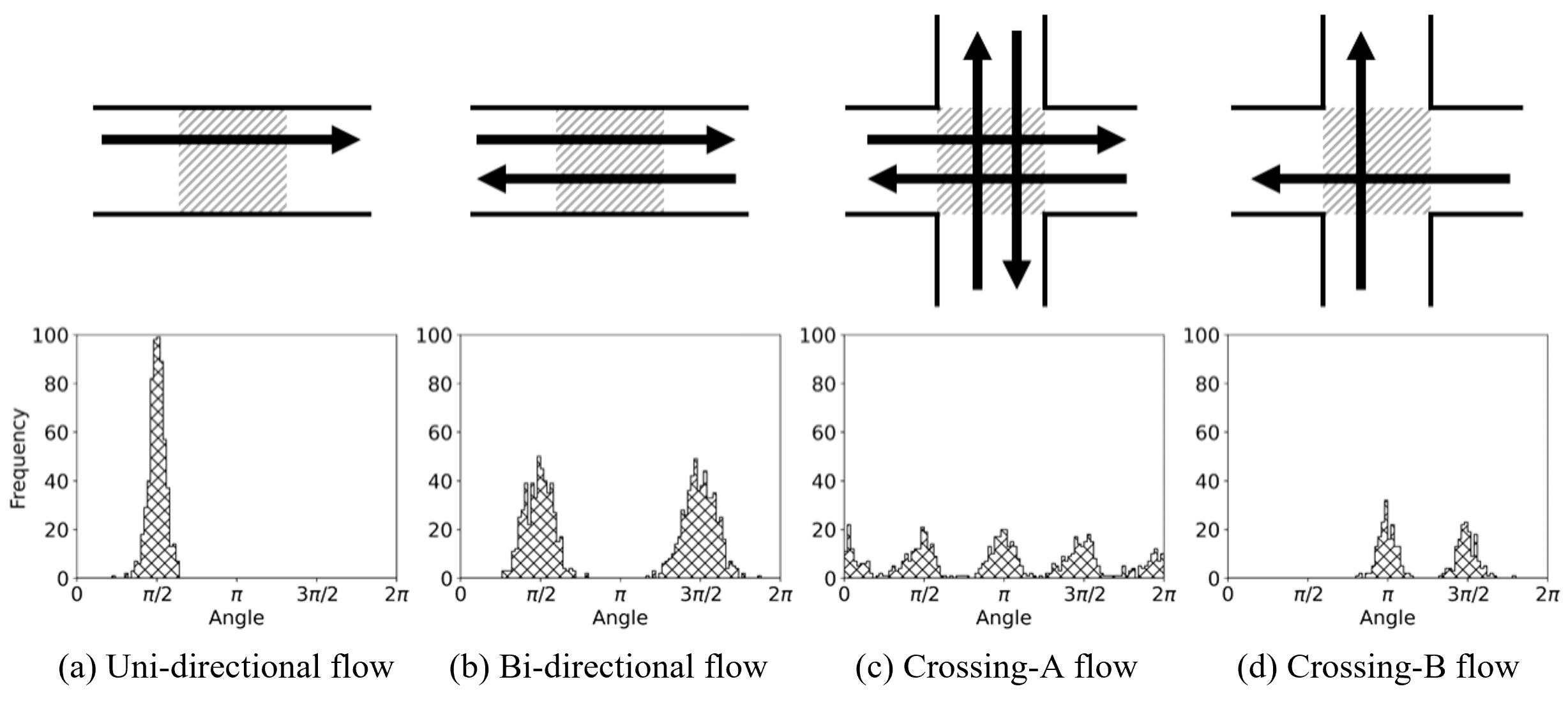}
    \caption{Four examples of flow types and their histograms. The data is a 10-second sample for each flow type from the data used in the case study.}
    \label{fig:EFT}
\end{figure}

Flow types could be quantitatively distinguished by the angle of the pedestrians movement direction.
The histograms in Figure \ref{fig:EFT} show the distributions of the pedestrians movement direction obtained from the actual data.
Uni/bi-directional flow means the flow passing through a corridor uni/bi-directionally.
Crossing-A flow is defined as the flow through a crossing bi-directionally, and crossing-B flow is defined as the flow through a crossing uni-directionally.
It is clear that the distribution shape of the angle is different for each flow type.
Therefore, at least in qualitatively, flow types can be classified by the shape of the distribution of the angle of the pedestrians movement direction.

\begin{figure}[tbp]
    \centering
    \includegraphics[width=0.8\hsize]{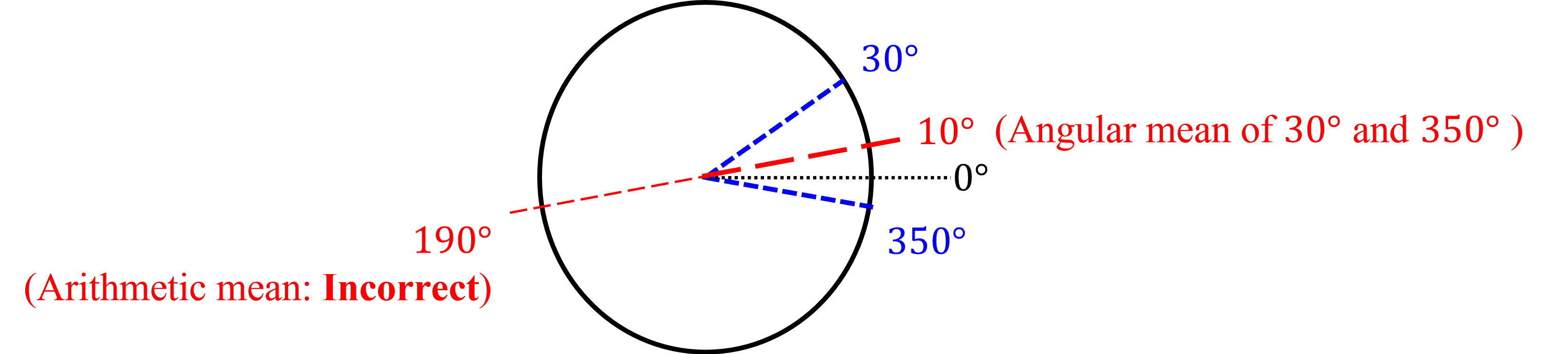}
    \caption{The example of angular mean and arithmetic mean for angle.}
    \label{fig:ang_exp}
\end{figure}

However, angles are difficult to analyze quantitatively or statistically due to its periodicity.
Let us consider an average of 30$^\circ$ and 350$^\circ$ angles shown in Figure \ref{fig:ang_exp}.
Clearly, the meaningful average should be 10$^\circ$.
However, the arithmetic mean gives 190$^\circ$, which is unreasonable.
This poses us significant difficulty for computing statistics of pedestrian moving direction such as the mean and variance of moving angles, which could be a useful statistics to characterize the flow type.

To tackle this challenge, this study employs {\it Directional Statistics} \cite{mardia2009directional}, which is a relatively new branch of Statistics specifically designed for analyzing angles.
It has been applied to several research fields such as wind direction \cite{johnson1978some} and animal movement direction \cite{schnute1992statistical}.
Recently, it is also applied to analyze transportation maps \cite{boeing2019urban}\cite{nagasaki2023understanding}\cite{nagasakiApplicationRoseDiagram2019}.
However, to the authors' knowledge, its application in dynamical transportation phenomena such as pedestrian flow is unseen.

This study aims to develop a pedestrian FD model that can be applied to various flow types regardless of its configuration.
The proposed model can capture the complicated pedestrian phenomena associated with flow types in a simple and straightforward manner by leveraging the Directional Statistics. 
To achieve this, we propose a novel statistic termed {\it $p$th angular variance} to effectively characterize pedestrian flow.
The features of the proposed model is validated by applying it to actual pedestrian trajectory data.
The contribution of this study can be summarized as follows.
\begin{itemize}
	\item We propose $p$th angular variance, a novel statistic based on Directional Statistics.
	\item We develop a pedestrian FD that uses angular distribution of pedestrian movement.
	\item To our knowledge, this work is the first application of Directional Statistics to dynamic transportation analysis.
\end{itemize}

\section{Literature review}

Pedestrian FDs have been studied widely in the literature.
The simplest studies on pedestrian FDs are those on uni-directional flows.
Based on data obtained from experiments or observations, many FD models of uni-directional flow have been proposed.

One of the earliest model is that proposed by Fruin \cite{fruin1970designing}. 
He formulated the flow $J$ as a function of density $\rho$ in the form of an equation as
\begin{gather}
    J = \frac{A\rho - B}{\rho^2},
\end{gather}
where $A$ and $B$ are parameters.
Virkler and Elayadath \cite{virkler1994pedestrian} applied several FD models that had been applied to vehicular flows to pedestrian flows.
Seyfried et al.~\cite{seyfried2005fundamental} showed from observation that a linear relationship between speed and the inverse of density can be established for single-row pedestrian flows.
Furthermore, they showed that the same relationship can be applied to general uni-directional flows by comparison with literature values.
Existing studies about uni-directional flows mostly modeled flow (or speed) as a function of density only.

Bi-directional flow is compared with uni-directional flow by pedestrian flow data in some studies.
Zhang and Seyfried \cite{zhang2013empirical} showed that the flow rate of bi-directional flow is significantly lower than that of uni-directional flow at high densities.
Lam et al.~\cite{lam2002study} showed that the capacity of a pedestrian space decreases in bi-directional flows.
These studies showed that bi-directional flow is less efficient than uni-directional flow.

On the other hand, an efficient condition of bi-directional flow, named lane formation, has also been studied \cite{feliciani2016empirical}\cite{jin2019observational}.
Jia et al.~\cite{jia2021pedestrian} proposed a parameter that represents the degree of deviation from the lane generated by lane formation and examined its correlation with the pedestrian's egress time from the exit.
The results showed that there was a positive correlation between the degree of deviation and the egress time, which suggests that the occurrence of lane formation makes the pedestrian flow more efficient.
Lee et al.~\cite{lee2016modeling} showed the capacity is improved by occurrence of lane formation in the simulation.

Multi-directional flows other than bi-directional flows have also been observed by researchers.
Zhang et al.~\cite{zhang2014comparison} compared uni-directional and T-junction flows and found that their FDs were different.
Cao et al.~\cite{cao2017fundamental} and Zhang and Seyfried \cite{zhang2013empirical} compared uni-directional, bi-directional, and crossing flows.
Both studies found that the capacities of bi-directional and crossing flows were smaller than those of uni-directional flows.
Iryo and Nagashima \cite{mihoPerformanceEvaluationPedestrian2015} showed that flows crossing angles of 45$^\circ$ and 135$^\circ$ tended to have lower flow than flows at 90$^\circ$, even at the same density.

Multi-directional flows are often described using physical models.
A well-known example is the social force model developed by Helbing and Molnar \cite{helbing1995social}.
In this model, the pedestrian is regarded as a point mass.
The motivation of the pedestrian moving to the desired direction, the influence between pedestrians, and the influence of walls are regarded as forces acting on the point mass.
The forced pedestrians' moves are based on the equations of motion. 
The social force model has been extended in various ways \cite{asano2009pedestrian}\cite{moussaid2009experimental} and has been successful in terms of reproducing the pedestrian flows.
Nakanishi and Fuse \cite{nakanishi2015preliminary} analyzed pedestrian movements by focusing on their moving direction.
They regarded pedestrian trajectories as data consisting of the pedestrian's position and angle of moving direction at each time.
A correlation between pedestrian position and direction of travel was discovered by basic analysis of pedestrian trajectories in front of ticket gates, and a regression model was constructed.
However, these studies were not performed on modeling pedestrian FD.

There are few studies modeling the FDs of pedestrian flows other than uni-directional flows.
Flötteröd and Lämmel \cite{flotterod2015bidirectional} formulated a parametric FD for bi-directional flow based on cellular automata.
Feliciani et al.~\cite{feliciani2018universal} modeled an FD for bi-directional flow by applying the flow ratio, which is defined as the ratio of flow in one direction and flow in the other direction.
However, these models cannot be applied to flow types other than bi-directional or uni-directional flow.
Moustaid and Flötteröd \cite{Moustaid2021pedestrian} proposed an FD model for multi-directional flow by extending a macroscopic node model \cite{flotterod2011node} for vehicular traffic flow.
Saberi and Mahmassani \cite{saberi2014exploring} applied Edie's definition \cite{edie} of fundamental traffic variables to the pedestrian trajectories to draw a FD for a multi-directional flow.
Wang et al.~\cite{wang2019fundamental} proposed new Voronoi diagram measurement method that takes into account the pedestrian's desired moving direction, and modelled a pedestrian FD applicable to multi-directional flow.
However, they requires pre-processing of pedestrian data to decompose the flow into specific number of directions.
It limits the generality of the model (i.e., an estimated model is specialized for a specific location with specific node-link configuration).
A model of pedestrian FDs comprehensively applicable to multiple flow types has not been developed.

\section{Model}\label{Model}

We formulate a pedestrian FD model that can capture the aforementioned complicated pedestrian dynamics by using directional statistics. 
The proposed model is formulated by extending a simple functional form of FD using the angular variance.

Note that in the following discussion, the unit of angle is radian.

\subsection{Angular variance}\label{sec:angvar}

First of all, we introduce an existing angular statistic termed {\it angular variance} as a basis of our model.
Angular variance~\cite{mardia2009directional} describes the degree of dispersion for angular data.
The angular variance $\nu_1$ for $N$ angular data $\theta_j~(j=1,2,\ldots,N)$ is defined as
\begin{align}
    \nu_1 = 1- \bar{R}_1, \label{eq1}
\end{align}
where
\begin{gather}
    \bar{R}_1 = \sqrt{\bar{C}_1^2 + \bar{S}_1^2}, \label{eq2} \\
    \bar{C}_1 = \frac{1}{N} \sum_{j=1}^{N} \cos \theta_j   , \quad \bar{S}_1 = \frac{1}{N} \sum_{j=1}^{N} \sin \theta_j. \label{eq3}
\end{gather}

\begin{figure}[tbp]
    \centering
    \includegraphics[width=0.8\hsize]{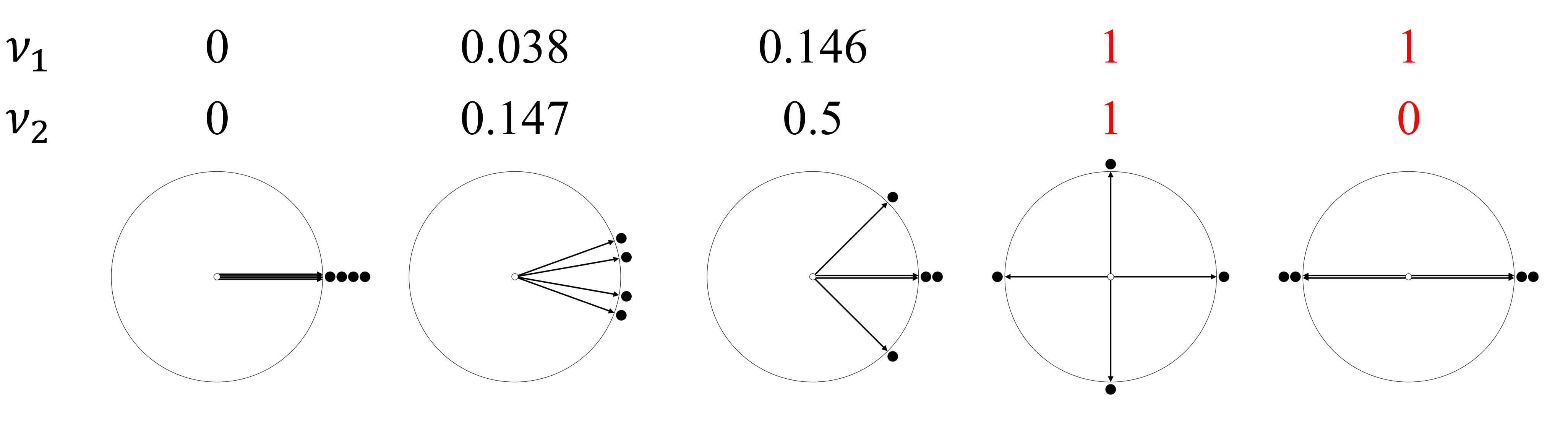}
    \caption{Angular data and corresponding angular variance $\nu_1$ and second angular variance $\nu_2$. The dots around the unit circle represent the angular data $\theta_j$ and the arrows represent the unit vector $(\cos \theta_j, \sin \theta_j)^{\top}$ corresponding to the data $\theta_j$.}
    \label{fig:angvar}
\end{figure}

The angular variance $\nu_1$ takes a value between $0$ and $1$. 
The value of ${\nu}_1$ gets larger when the data is scattered, which is consistent with the conventional statistics.
Figure \ref{fig:angvar} shows examples of angular data and variance $\nu_1$.
As shown in the left of the figure, when the angular data are concentrated in a close direction, $\nu_1$ takes a value close to 0. 
On the other hand, when the angular data are dispersed in opposite directions, $\nu_1$ takes a value close to 1.

Technical details behind the definition of angular variance is as follows.
For the $j$th angular data $\theta_j$, $(\cos \theta_j, \sin \theta_j)^{\top}$ represents the unit vector for that angle, where $^\top$ means the transpose of vector. 
Therefore, $(\bar{C}_1, \bar{S}_1)^{\top}$ is the composite of $N$ unit vectors divided by its data size $N$.
$\bar{R}_1$ is the norm of $(\bar{C}_1, \bar{S}_1)^{\top}$, which takes a value between $0$ and $1$, and it varies depending on the degree of dispersion of the data.
Because of this definition, the angular variance can be computed for any kind of angular data regardless of the periodicity issue.

\subsection{$p$th angular variance}

The angular variance is useful to quantify the degree of dispersion of angular data.
However, as shown on the right of Figure \ref{fig:angvar}, the data with multiple peaks equally distributed always has $\nu_1=1$.
This means that the data in Figures \ref{fig:EFT}(b) and (c) cannot be distinguished if only $\nu_1$ is used.
We need new statistic to distinguish such data with different number of multiple peaks.

Here, we propose a novel statistic {\it $p$th angular variance} $\nu_p~(p\in\mathbb{N})$, defined as 
\begin{gather}
    \nu_p = 1- \bar{R}_p \label{eq4},
\end{gather}
where
\begin{gather}
    \bar{R}_p = \sqrt{\bar{C}_p^2 + \bar{S}_p^2}, \label{eq5} \\
    \bar{C}_p = \frac{1}{N} \sum_{j=1}^{N} \cos p \theta_j   , \quad \bar{S}_p = \frac{1}{N} \sum_{j=1}^{N} \sin p \theta_j. \label{eq6}
\end{gather}
$p$ is an exogenous variable that can take any natural number.
The $p$th angular variance is a generalization the original angular variance, and is applicable to distinguish whether data has $p$ peaks or not.

The $p$th angular variance $\nu_p$ takes a value between $0$ and $1$, similar to the original angular variance.
The value of $\nu_p$ gets smaller when the data has $p$ evenly-distributed peaks; otherwise it gets larger.
Contrary, the value of $\nu_m$ for $m \neq p$ usually get lager.
Therefore, $\nu_p$ can be used to quantify the number of peaks in angular data.
These properties will be formalized later in Theorem \ref{thm1}.

Figure \ref{fig:angvar} also shows examples of second angular data and variance $\nu_2$.
The second angular variance with the two peaks shown on the far right is 0.
On the other hand, the second angular variance remains 1 with four peaks in the second from the right.
The data with two peaks and four peaks, which were indistinguishable with the conventional angular variance, are distinguishable with the second angular variance.

As described above, the $p$th angular variance has a favorable property as a statistic to distinguish data with multiple peaks, because $\nu_p$ for such data is significantly different depending on $p$.
This property can be formalized as the following theorem

\begin{thm}\label{thm1}
Let $\nu_p(\bm{\theta})$ denote the $p$th angular variance for a dataset $\bm{\theta}=\{\theta_j\,|\,j=1,2,\dots,N\}$ and $\tilde{\bm{\theta}}^m$ denote a dataset with $2\pi/m~(m\in\mathbb{N}, m\geq2)$ period, the following formulas hold

\begin{align*}
\nu_p(\tilde{\bm{\theta}}^m)=1\qquad & \text{if $p\leq{m-1}$,} \\
0\leq\nu_p(\tilde{\bm{\theta}}^m)\leq1\qquad & \text{if $p=m$.}    
\end{align*}
\end{thm}
\noindent\textbf{Proof.} See \ref{prooftheo}.
\\
\\
\noindent Because of these theoretical properties, the $p$th angular variance $\nu_p$ is useful for quantifying the angular data with $2\pi/m$ period (e.g., angular data with multiple peaks).

\begin{figure}[tbp]
    \centering
    \includegraphics[width=14cm]{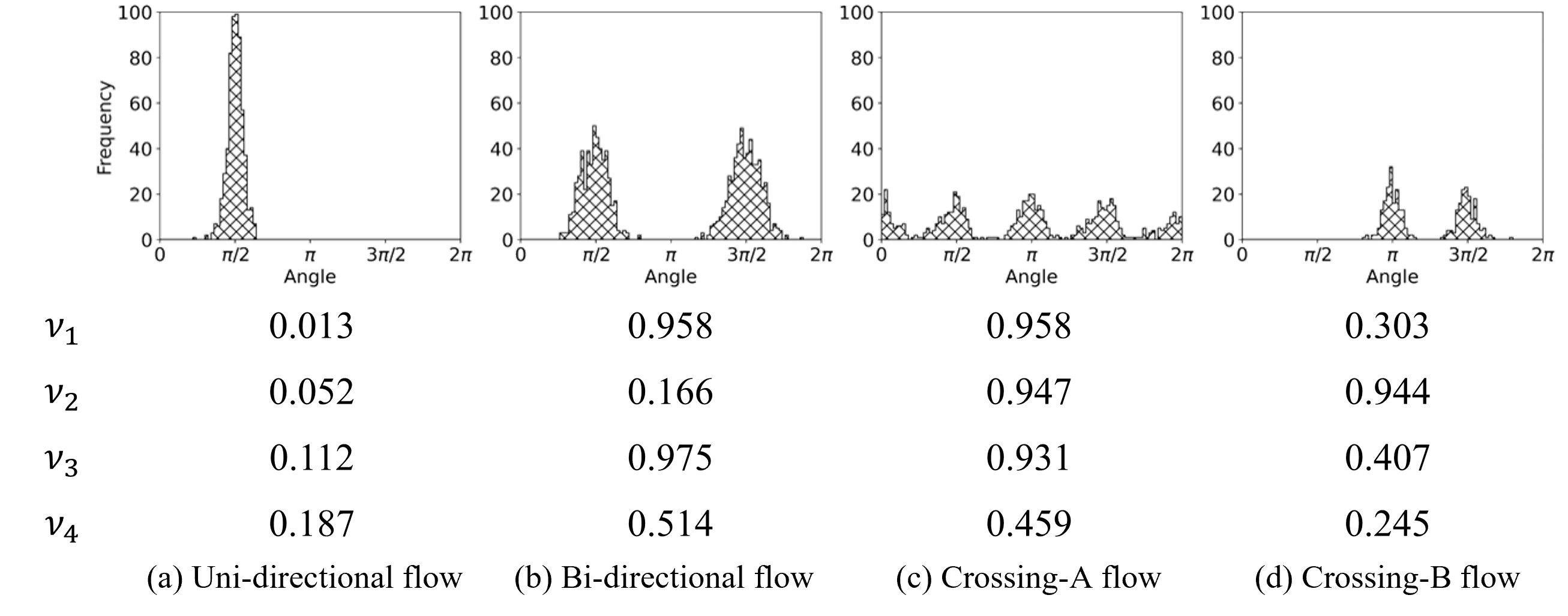}
    \caption{Histograms for each flow type and corresponding values for each $p$th angular variance.}
    \label{fig:EFT2}
\end{figure}

Figure \ref{fig:EFT2} shows pedestrian flow's histograms for each flow type and corresponding values for each $p$th angular variance $\nu_p~(p=1,2,3,4)$.
The angular variance $\nu_1$ of bi-directional flow and crossing-A flow takes values close to 1, this means $\nu_1$ cannot distinguish these flows solely.
In contrast, the second angular variance $\nu_2$ of bi-directional flow and the fourth angular variance $\nu_4$ of crossing-A flow take small values.
Therefore, these flows can be distinguished by $\nu_2$ and $\nu_4$.
In addition, $\nu_p$ for all $p$ of the uni-directional flow take a value close to 0.
$\nu_1$ of crossing-B flow has a relatively smaller value than bi-directional flow and crossing-A flow because the interval between peaks is narrow, while $\nu_2$ takes a larger value.

The four flow types can be distinguished by a combination of $\nu_1$ and $\nu_2$ without using a higher degree of angular variances.
The uni-directional flow has small angular variance $\nu_1$ and a small second angular variance $\nu_2$, bi-directional flow has large $\nu_1$ and small $\nu_2$, crossing-A flow has large $\nu_1$ and large $\nu_2$ and crossing-B flow has relatively small $\nu_1$ and large $\nu_2$. 
Therefore, the four flow types can be distinguished by the angular variance $\nu_1$ and the second angular variance $\nu_2$.

\subsection{Formulation of FD}

Now we formulate the proposed FD by utilizing the aforementioned angular statistics.
Note that in the following discussion, $J$ is the flow, $u$ is the free flow speed, $\rho$ is the density and $C$ is the capacity.

\subsubsection{Base function}

As a basis of the proposed model, a FD function without angular effects is formulated first based on the following common assumptions.
In free flow state with $u\rho \leq C$, the flow $J$ equals $u\rho$ and increases linearly with the density $\rho$.
In congested state with $u\rho>C$, the flow $J$ equals the capacity $C$ which is constant to simplify the model.
Under these conditions, the simple pedestrian FD models can be described as
\begin{gather}
    J=\min(u \rho, C). \label{eq7}
\end{gather}

In this study, the flow $J$ is assumed to be constant in the congested state independent of the density $\rho$.
In general, the flow may decrease as the density increases in the congested regime (c.f., the triangular FD).
Note that it is very easy to extend the proposed model (\ref{eq7}) to accommodate this phenomenon; see Appendix \ref{ap2} for a case study.
However, our case study is revealed that, at least for the particular dataset (see Fig.~\ref{fig:DF}), the flow did not decrease in a statistically significantly manner.
In fact, this is a common phenomenon in multi-directional pedestrian flows; unless the density was dangerously high, flow does not significantly decrease in the congested regime \cite{virkler1994pedestrian}\cite{zhang2014comparison}\cite{cao2017fundamental}\cite{flotterod2015bidirectional}.
Therefore, this assumption on constant flow in congested regime can be justified and easy to be generalized if necessary.

\subsubsection{Incorporating $p$th angular variance into FD function}\label{sec:angle}

In the following, angular variance is incorporated into the model to represent the effects of the avoidance of conflicts on pedestrian flows. 
When pedestrian flows cross, the flow decreases due to the avoidance.
The avoidance is expected to occur primarily at high density, i.e., in congested state.
Therefore, the avoidance of conflicts is assumed to affect capacity $C$. 

First, the effect of avoidance of conflicts is modeled.
If pedestrians flow uni-directional, the effect seems to be small.
In this case, the distribution of the direction of pedestrians has a single peak.
On the other hand, if pedestrians flow multi-directional, the effect seems to be large.
In this case, the distribution of the direction of pedestrians does not have a single peak clearly.
As described in section \ref{sec:angvar}, the degree of dispersion in the direction of the pedestrian can be described by angular variance $\nu_1$. 
Therefore, the magnitude of the capacity $C$ reduction due to avoidance of conflicts is controlled by angular variance $\nu_1$ in the proposed model.

Next, the effect of lane formation is modeled.
Lane formation occurs in bi-directional flow.
For bi-directional flow, the second angular variance $\nu_2$ takes a small value because the distribution of the direction of pedestrians has two peaks with a period of $\pi$.
Since the capacity $C$ is expected to decrease when lane formation does not occur, the capacity $C$ decreases as the second angular variance $\nu_2$.

\subsubsection{Incorporating the effects of the walls into FD function}\label{sec:wall}

\begin{figure}[tbp]
    \centering
    \includegraphics[width=0.6\hsize]{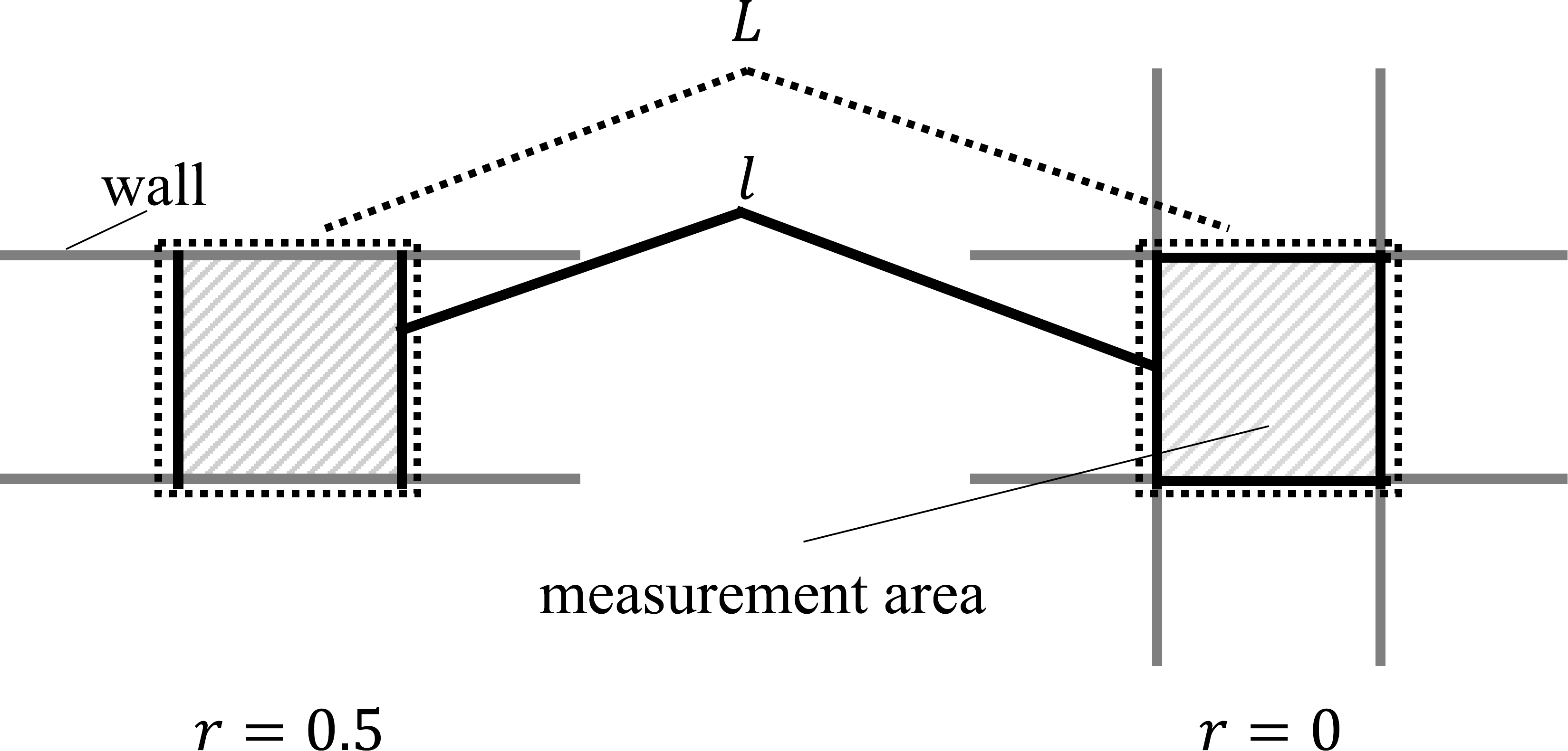}
    \caption{Examples of wall ratio $r$.}
    \label{fig:WR}
\end{figure}

Additionally, the model incorporates the effect of surrounding walls within the observation area. 
When walls are placed in the travel area, pedestrians move to avoid conflicting walls.
Therefore, movement in an area surrounded by walls is expected to be less efficient.
To quantify the effect of walls, we defined a metric termed the \textit{wall ratio} $r$ as,
\begin{gather}
    r=\frac{L-l}{L}, \label{eqwr}
\end{gather}
where $L$ denotes the total perimeter of the measurement area and $l$ denotes the length of the passable pedestrian section within it. 
For example, as shown in Figure \ref{fig:WR} when square measurement areas are assumed, $r$ equals 0.5 for a corridor and $r$ equals 0 for a crossing. In the model, the capacity $C$ is assumed to decrease with the wall ratio $r$ increasing.

\subsubsection{Proposed FD model}
Based on the discussion in section \ref{sec:angle} and \ref{sec:wall}, the capacity $C$ is assumed to decrease due to angular variance $\nu_1$, the second angular variance $\nu_2$, and wall ratio $r$.
Therefore, the capacity is expressed as
\begin{align}
    C=C_0(1-\gamma_1\nu_1)(1-\gamma_2\nu_2)(1-\gamma_\mathrm{wall}r), \label{eq11}
\end{align}
where $C_0$ represents the capacity without the effects of $\nu_1$, $\nu_2$ and $r$. $\gamma_1,\gamma_2$ and $\gamma_\mathrm{wall}$ are parameters.
From now on, the term of $(1-\gamma_1\nu_1)(1-\gamma_2\nu_2)(1-\gamma_\mathrm{wall}r)$ is called \textit{capacity penalty}.

Substituting Eq.~(\ref{eq11}) into Eq.~(\ref{eq7}), an FD model of the pedestrian flows that takes into account the effects of flow types and the wall is obtained as
\begin{gather}
    J=\min(u \rho, C_0(1-\gamma_1\nu_1)(1-\gamma_2\nu_2)(1-\gamma_\mathrm{wall}r)).\label{eqx}
\end{gather}

Since Eq.~(\ref{eqx}) uses the minimum function, there are indifferentiable points.
To simplify handling in numerical processing such as estimation, Eq.~(\ref{eqx}) is smoothed by applying the log-sum-exp (LSE) function \cite{nielsen2016guaranteed} defined as
\begin{align}
    \mathrm{LSE}(x_1,x_2,\ldots,x_n) &\equiv \log \left( \sum_{j=1}^{i} \exp(x_i) \right) \nonumber \\
    &\thickapprox \max (x_1,x_2,\ldots,x_n). \label{eq8}
\end{align}
LSE function is a smooth approximation of the maximum function. 
It can also approximate the minimum function by interchanging the signs as
\begin{align}
    -\mathrm{LSE}(-x_1,-x_2,\ldots,-x_n) &\thickapprox -\max (-x_1,-x_2,\ldots,-x_n) \nonumber \\
    &= \min (x_1,x_2,\ldots,x_n). \label{eqxx}
\end{align}

The pedestrian FD model proposed in this study is completed by applying the LSE function with the signs switched to Eq.~(\ref{eqx}) as
\begin{align}
    J(\rho,\nu_1, \nu_2, r)=-\log \{\exp(-u\rho) + \exp\left(-C_0(1-\gamma_1\nu_1)(1-\gamma_2\nu_2)(1-\gamma_\mathrm{wall}r)\right)\}. \label{eq12}
\end{align}

\begin{figure}[tbp]
    \centering
    \includegraphics[width=0.7\hsize]{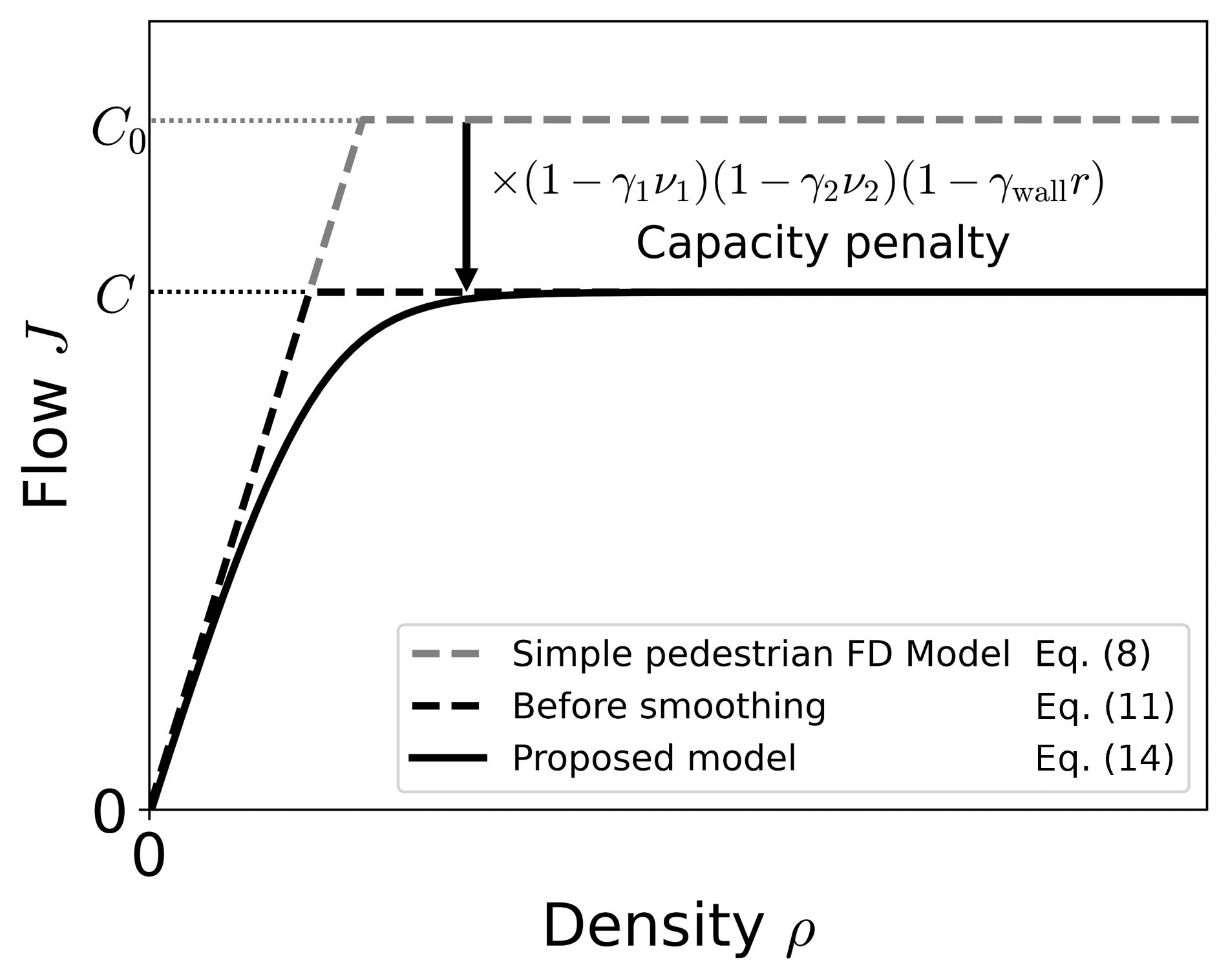}
    \caption{Schematic diagram of the models.}
    \label{fig:MZ}
\end{figure}

Figure \ref{fig:MZ} shows a schematic diagram of the models described in this section.

\section{Case study}

We validate the proposed model using actual pedestrian flow datasets.
The validation is conducted by parameter significance test and fitting analyses (section \ref{sec_parameter_estimation}), comparison with other models (section \ref{sec_ablation}), and investigating the detailed features (section \ref{sec_est_discussion}).

\subsection{Data}
We used “Unidirectional pedestrian flow in a corridor” dataset (DOI: 10.34735/ped.2013.6), “Bidirectional pedestrian flow in a corridor” dataset (DOI: 10.34735/ped.2013.5) and “Pedestrian flow at a 90 degree crossing” dataset (DOI: 10.34735/ped.2013.4) published by Forschungszentrum Jülich.
These three datasets correspond to four flow types: uni-directional flow, bi-directional flow, crossing-A flow, and crossing-B flow, defined in Section \ref{chap:intro}.
The data contains the pedestrian's coordinates and ID for each frame.
The frame rates are 25 fps or 16 fps, depending on the dataset.

\begin{figure}[tbp]
    \centering
    \includegraphics[width=0.5\hsize]{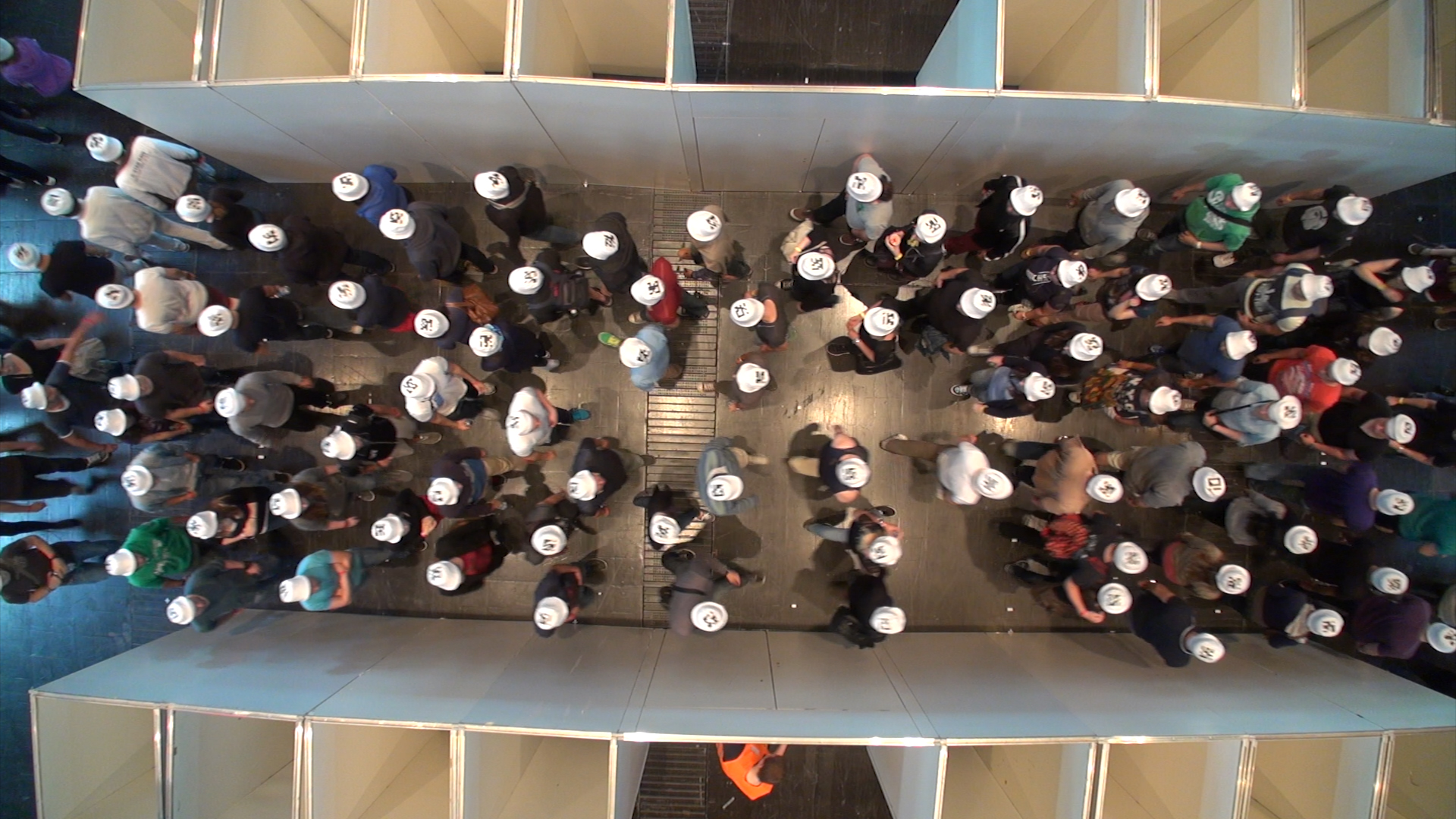}
    \caption{Top view photograph of one of the bi-directional flow experiment (Source: \protect\url{https://ped.fz-juelich.de/da/doku.php?id=corridor6}).}
    \label{fig:experiment_photo}
\end{figure}

The flow $J$ and the density $\rho$ are calculated from the trajectory data. They were obtained by applying the Edie's definition \cite{edie} as
\begin{gather}
    J=\frac{\sum_s d_s}{A}, \label{eq13}
\end{gather}
and
\begin{gather}
    \rho=\frac{\sum_s t_s}{A}, \label{eq14}
\end{gather}
respectively.
$A$ is the spatio-temporal region in which the flow and density are measured.
The time region was set to 10 seconds, and the spatial region was set to square areas in the center of the corridor or crossing. $d_s$ represents the distance that pedestrian $s$ has traveled in spatio-temporal region $A$, and $t_s$ represents the travel time in the region.
The flow $J$ and the density $\rho$ at time $\tau$ were calculated in the following way.

\begin{enumerate}
    \item The coordinates of the pedestrians in the measurement area at time $\tau$ are obtained from the data.
    \item The number of pedestrians in the measurement area multiplied by the interval for calculating (1 second) was added to $\sum_s t_s$.
    \item The displacements for all pedestrians in the measurement area were calculated from the coordinates of the pedestrians after 1 second. They were added to $\sum_s d_s$.
    \item Steps 1 to 3 were repeated for 10 seconds while increasing the time by 1 second.
    \item The flow $J$ and density $\rho$ were calculated by Eq.~(\ref{eq13}) and Eq.~(\ref{eq14})
\end{enumerate}

The interval for calculating the displacement was set as long as 1 second. 
By taking a long interval, short movements in unintended directions due to the avoidance of conflicts were not added to the flow.

\begin{figure}[tbp]
    \centering
    \includegraphics[width=0.8\hsize]{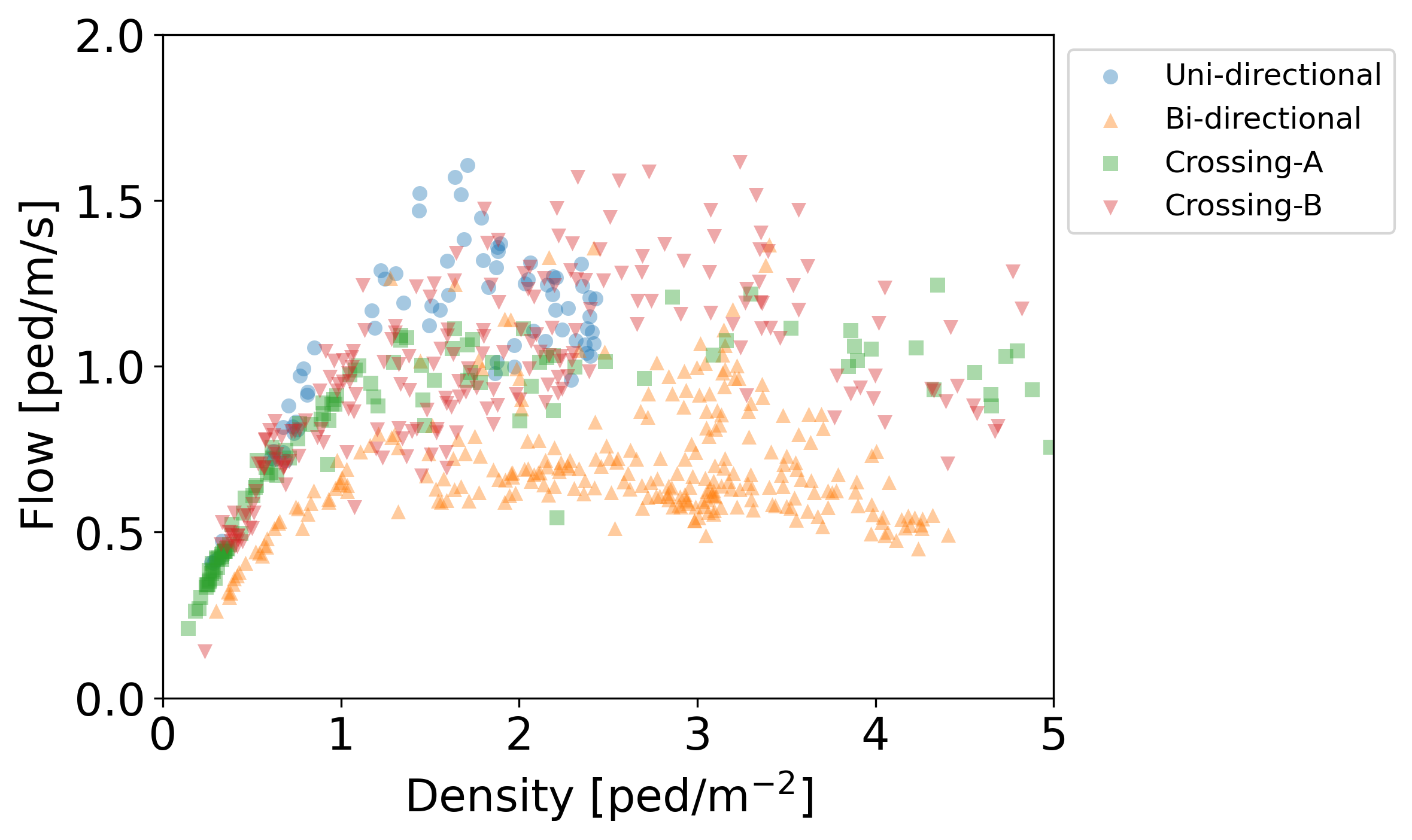}
    \caption{Flow-density plots of the pedestrian datasets.}
    \label{fig:DF}
\end{figure}

Figure \ref{fig:DF} shows a scatter plot of the calculated flow $J$ and density $\rho$ for four flow types. The capacity of the FD is different depending on the flow types. Therefore, explaining this data with a single equation is difficult with existing models.

The angular variance $\nu_1$ and the second angular variance $\nu_2$ at time $\tau$ were calculated in the following way. 

\begin{enumerate}
    \item The coordinates of the pedestrians in the measurement area at time $\tau$ are obtained from the data.
    \item Angles of pedestrians were calculated by comparing with the coordinates after 0.2 (for 25 fps data) or 0.25 (for 16 fps data) seconds.
    \item Steps 1 to 2 were repeated for 10 seconds while increasing the time by 0.2 or 0.25 seconds.
    \item $\nu_1$ and $\nu_2$ were calculated using all the angles computed during the 10 seconds by Eq.~(\ref{eq1}) and Eq.~(\ref{eq4}).
\end{enumerate}
In order to evaluate the avoidance of conflicts by angular variance, the intervals were set to short enough to pick up small movements.

The wall ratio $r$ was calculated according to the definition given in Eq.~(\ref{eqwr}), with $r=0.5$ for uni-directional and bi-directional flows and $r=0$ for crossing flows.

The flow $J$, the density $\rho$, the angular variance $\nu_1$, the second angular variance $\nu_2$, and the wall ratio $r$ were calculated every 10 seconds according to the above procedure.
For data cleansing, remove the first 10 seconds and the last 10 seconds of data of the experiment.
Finally, 70 sets of the variables at random time intervals were sampled, of which 40 were set as training data and 30 as test data for each flow type.

\subsection{Parameter estimation}\label{sec_parameter_estimation}
The parameters of the model were estimated using the training data. 
The data for all flow types were applied simultaneously to obtain the parameters that fit all flow types.
The least squares method was employed for the estimation. 
Table \ref{suitei} shows the parameter estimation results.
Note that the dimensions for parameters are as follows---free flow speed $u$: m/s, capacity $C_0$: ped/m/s, coefficient of reduction in flow by angular variance $\gamma_1$: dimensionless, coefficient of reduction in flow by second angular variance $\gamma_2$: dimensionless, and coefficient of reduction in flow by wall $\gamma_\mathrm{wall}$: dimensionless.

\begin{table}[htb]
    \centering
    \caption{Estimation results.}
    \label{suitei}
    \begin{tabular}{c|crc}
    \hline
                         & Coefficient   & $t$ value & $p$ value              \\ \hline
    $u$      & 3.262  & 16.873 & $<10^{-10}$       \\
    $C_0$ & 1.566  & 11.892 & $<10^{-10}$       \\
    $\gamma_1$           & 0.266 & 7.327 & $<10^{-10}$ \\
    $\gamma_2$           & 0.221 & 2.655 & $8.765\times 10^{-3}$ \\
    $\gamma_{\mathrm{wall}}$      & 0.486 & 4.470 & $1.510\times 10^{-5}$ \\ \hline
    \end{tabular}
\end{table}

According to the estimation results in Table \ref{suitei}, the signs of the estimates of $\gamma_1$, $\gamma_2$, and $\gamma_\mathrm{wall}$ were positive.
This is consistent with the assumption in Section \ref{Model} that the larger angular variance $\nu_1$, the second angular variance $\nu_2$, and the wall ratio $r$ decrease the capacity $C$.
In addition, all parameters were significant according to the $t$ values and $p$ values.

Table \ref{R2} shows the value of the coefficients of determination $R^2$ and adjusted coefficients of determination $\bar{R}^2$ for the training and test data.
The coefficient of determination was approximately 0.6, indicating that this model can explain flow $J$ of different flow types to some extent.

\begin{table}[htb]
    \centering
    \caption{Coefficient of determination of the proposed model.}
    \label{R2}
    \begin{tabular}{c|cc}
    \hline
           & $R^2$ & $\bar{R}^2$ \\ \hline
    Train data  & 0.663 & 0.653       \\
    Test data & 0.713 & 0.701     \\ \hline
    \end{tabular}
\end{table}

\subsection{Ablation study}\label{sec_ablation}
We estimated a model without $\nu_1$ and $\nu_2$ that corresponds to Eq.\ (\ref{eq9}) and without $\nu_2$ that corresponds to Eq.\ (\ref{eq10}) to determine how much they contributed to the performance of the model.
\begin{align}
    J=-\log \{ \exp(-u\rho) + \exp\left(-C_0(1-\gamma_\mathrm{wall}r)\right)\}. \label{eq9}
\end{align}
\begin{align}
    J=-\log \{ \exp(-u\rho) + \exp\left(-C_0(1-\gamma_1\nu_1)(1-\gamma_\mathrm{wall}r)\right)\}. \label{eq10}
\end{align}
From now on, the model Eq.~(\ref{eq9}) and Eq.~(\ref{eq10}) is called \textit{base model} and \textit{$\nu_1$ model}, respectively.

Table \ref{suitei2} and Table \ref{suitei3} show the parameter estimation results and Table \ref{R22} and Table \ref{R23} shows the value of the coefficients of determination $R^2$ and adjusted coefficients of determination $\bar{R}^2$ for the training and test data.
The speed parameter $u$, which represents the gradient in the free flow state,is not much different from that in Table \ref{suitei} for both base model and $\nu_1$ model.
On the other hand, the capacity parameter $C_0$, which represents the capacity in the congested state, has a smaller value than that in Table \ref{suitei}.
This means that the performance of the pedestrian space is underestimated without taking into account the effect of angular variances.
In addition, a comparison of Table \ref{R2} and Table \ref{R22} and Table \ref{R23} shows that the accuracy of the estimation naturally decreases when FDs of different flow types are estimated simultaneously without the incorporation of the angular variance terms.

\begin{table}[htb]
    \centering
    \caption{Estimation results of the base model.}
    \label{suitei2}
    \begin{tabular}{c|crc}
    \hline
                         & Coefficient   & $t$ value & $p$ value              \\ \hline
    $u$         & 3.674  & 12.709 & $<10^{-10}$       \\
    $C_0$ & 1.020  & 32.655 & $<10^{-10}$       \\
    $\gamma_{\mathrm{wall}}$      & 0.134 & 1.851 & $0.066$ \\ \hline
    \end{tabular}
\end{table}

\begin{table}[htb]
    \centering
    \caption{Estimation results of the $\nu_1$ model.}
    \label{suitei3}
    \begin{tabular}{c|crc}
    \hline
                         & Coefficient   & $t$ value & $p$ value              \\ \hline
    $u$         & 3.369  & 16.533 & $<10^{-10}$       \\
    $C_0$ & 1.301  & 29.029 & $<10^{-10}$       \\
    $\gamma_1$ & 0.314  & 11.189 & $<10^{-10}$       \\
    $\gamma_{\mathrm{wall}}$ & 0.243 & 4.298 & $3.037\times 10^{-5}$ \\ \hline
    \end{tabular}
\end{table}

\begin{table}[htb]
    \centering
    \caption{Coefficient of determination of the base model.}
    \label{R22}
    \begin{tabular}{c|cc}
    \hline
           & $R^2$ & $\bar{R}^2$ \\ \hline
    Train data  & 0.444 & 0.409       \\
    Test data & 0.433 & 0.394       \\ \hline
    \end{tabular}
\end{table}

\begin{table}[htb]
    \centering
    \caption{Coefficient of determination of the $\nu_1$ model.}
    \label{R23}
    \begin{tabular}{c|cc}
    \hline
           & $R^2$ & $\bar{R}^2$ \\ \hline
    Train data & 0.647 & 0.638       \\
    Test data & 0.678 & 0.667       \\ \hline
    \end{tabular}
\end{table}

\subsection{Discussion}\label{sec_est_discussion}

\begin{figure}[tbp]
    \centering
    \includegraphics[width=0.8\hsize]{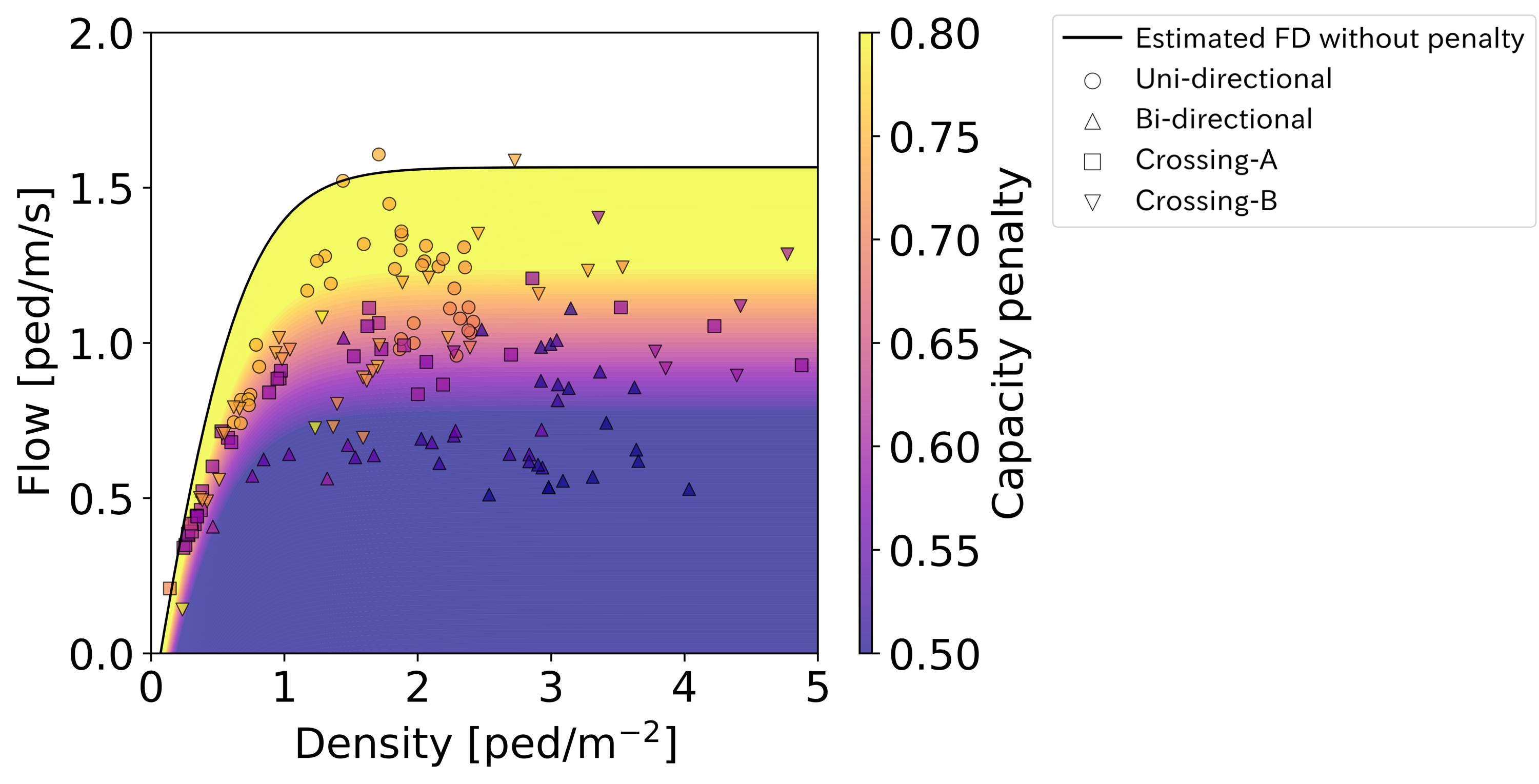}
    \caption{Flow-density plots of (the 40 training data) $\times$ (4 flow types) and the estimated FD colored according to the value of capacity penalty $(1-\gamma_1\nu_1)(1-\gamma_2\nu_2)(1-\gamma_\mathrm{wall}r)$.}
    \label{fig:FDC}
\end{figure}

Now we discuss the features of the model by investigating the detailed results.
Figure \ref{fig:FDC} shows the estimated FD and plots of samples on the flow-density plane.
The FD drawn as the solid line is the case without any capacity penalty where the angular variance $\nu_1$, the second angular variance $\nu_2$, and wall ratio $r$ are all $0$, i.e., the capacity $C$ equals $C_0$.
As the capacity penalty increases, the FD moves down from the solid line.
The domain is colored according to the magnitude of the capacity penalty.
On the other hand, the dots indicating the samples with the flow and density for a particular 10-second period are also colored by their capacity penalty.
Therefore, when the color of a dot and the surrounding area are similar, it implies a favorable estimation result for that sample.

According to Figure \ref{fig:FDC}, most light-colored dots with a small penalty are located in the light domain, and most dark-colored dots with a large penalty are located in the dark domain.
In addition, the results for each flow type identified by the dots' shapes also show a similar color match between the dots and the surrounding domain for all four types.
This indicates that the proposed model performs well for data with a mixture of several flow types.

\begin{figure}[tbp]
    \centering
    \includegraphics[width=\hsize]{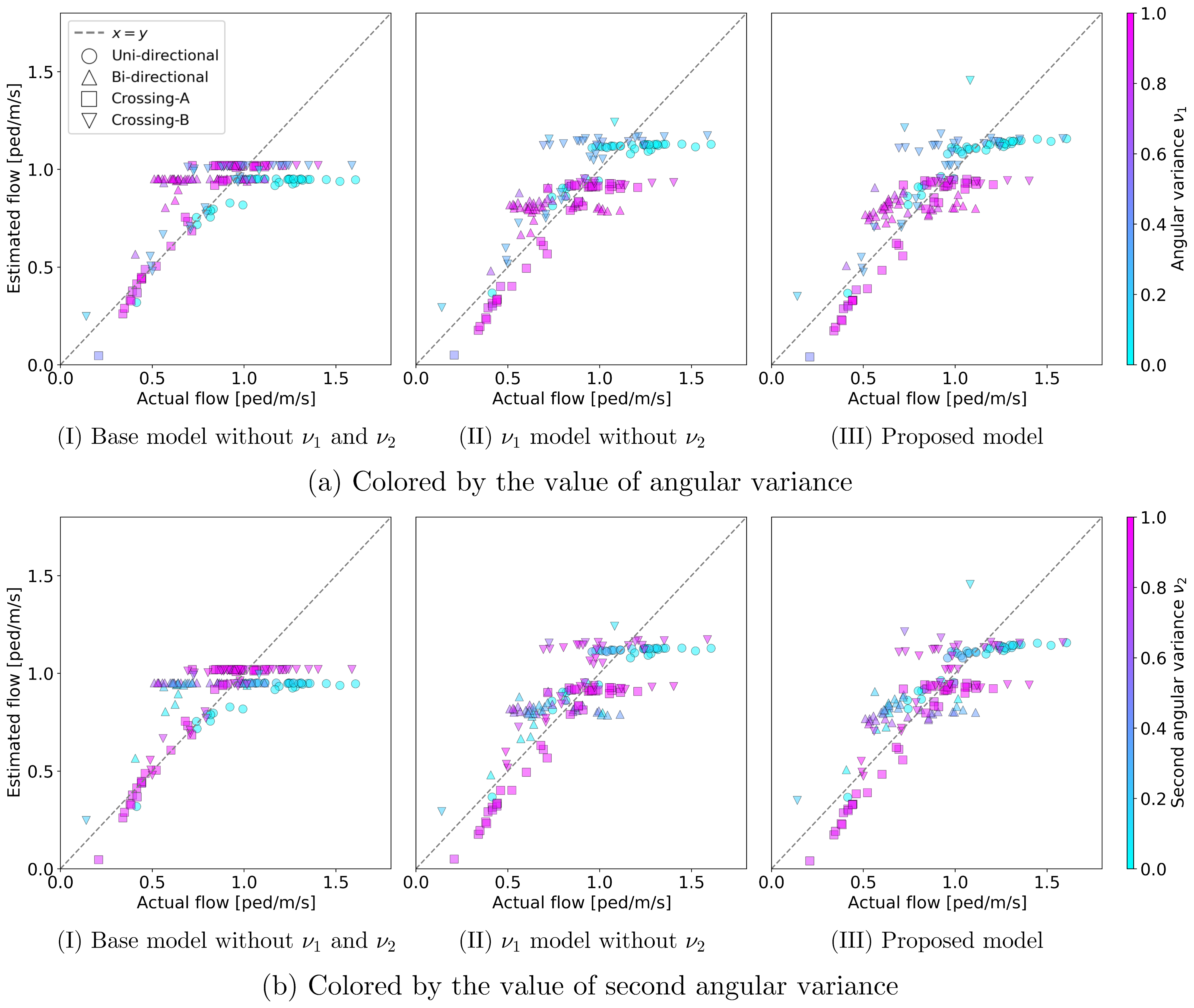}
    \caption{Scatter plots with actual flow on the horizontal axis and the estimated flow on the vertical axis for each model with colored by the value of angular variance $\nu_1$ or that of second angular variance $\nu_2$.}
    \label{fig:actest}
\end{figure}

Figure \ref{fig:actest} compares the estimated flow and the actual flow for each sample.
The left figure shows the comparison with the base model, the middle figure with the $\nu_1$ model, and the right figure with the proposed model.
Each sample is colored by the value of angular variance $\nu_1$ in Figure \ref{fig:actest}(a) and the one of second angular variance $\nu_2$ in Figure \ref{fig:actest}(b).

Flows estimated to be close to $1.0$ ped/m/s in the base model are separated by incorporating $\nu_1$.
In concrete, uni-directional flows with an actual flow distributed between $1.0$ and $1.5$ colored by blue in both figures are separated from the other flow types.
The flow is estimated to be larger in the uni-directional flow because the angular variance $\nu_1$ and the second angular variance $\nu_2$ are smaller.
Furthermore, the bi-directional flow drawn by the triangular dots locates around $(\text{Actual flow}, \text{Estimated flow})=(0.6, 0.8)$ in Figure \ref{fig:FDC}(b)(I\hspace{-1.2pt}I) are vertically distributed according to the value of $\nu_2$ in the proposed model, resulting in an improvement in the accuracy of the fit.
This implies that the lane formation of the bi-directional flow can be represented to a certain extent by the second angular variance $\nu_2$.
These indicate that incorporating the angular variance $\nu_1$ and the second angular variance $\nu_2$ contribute to simultaneously estimating the pedestrian FD of several flow types.

\section{Conclusion}

We proposed a comprehensive model of pedestrian FDs that can be applied to various flow types. 
Specifically, we use angular distribution of pedestrian movement by utilizing Directional Statistics, namely the angular variance (existing statistic) and the $p$th angular variance (novel statistic proposed by this study), to characterize the flow types in a universal manner.
It enables the FDs to explain the effect of flow types on flow performance, such as efficient bi-directional flows and inefficient crossing flows. 
The model was estimated and validated using actual pedestrian trajectory data. 
The estimated values of the parameters were consistent with the assumptions of the modeling. 
The capacities of estimated FD varied based on the effect of avoidance of conflicts, measured by angular variance and the second angular variance.
This implies that pedestrian flow characteristics can be depicted using Directional Statistics.

The current model is the first attempt of this kind and has spaces for improvements.
First, the proposed model does not describe the congested part of an FD.
This is because congested pedestrian flow is completely different from free-flowing or saturated flows, and experimental data is very limited.
Second, although it predict average flow based on directional-dependent flow, it does not predict direction-dependent flow itself (i.e., input is directional-dependent, but the output is not).
To extend the proposed model to directional-dependent flow prediction, the approach developed by Nagasaki and Seo \cite{nagasaki2023understanding} could be utilized.
Case studies with other datasets with different geometries such as in T-junctions or open spaces are also considerable to confirm the generality of the proposed methodology.

\appendix
\section{Proof of Theorem and Lemmas}\label{ap1}

\subsection{Proof of Theorem \ref{thm1}}\label{prooftheo}

Theorem \ref{thm1} is proved by the following two lemmas.

\begin{lem}\label{lem1}
$\nu_p(\tilde{\bm{\theta}}^m)$ is always equal to 1 when $p\leq{m-1}$.
\end{lem}
\noindent\textbf{Proof.} See \ref{proof1}.

\begin{lem}\label{lem2}
$\nu_m(\tilde{\bm{\theta}}^m)$ is not limited to 1.
\end{lem}
\noindent\textbf{Proof.} See \ref{proof2}.

\subsection{Proof of Lemma \ref{lem1}}\label{proof1}
\begin{figure}[tbp]
    \centering
    \includegraphics[width=0.6\hsize]{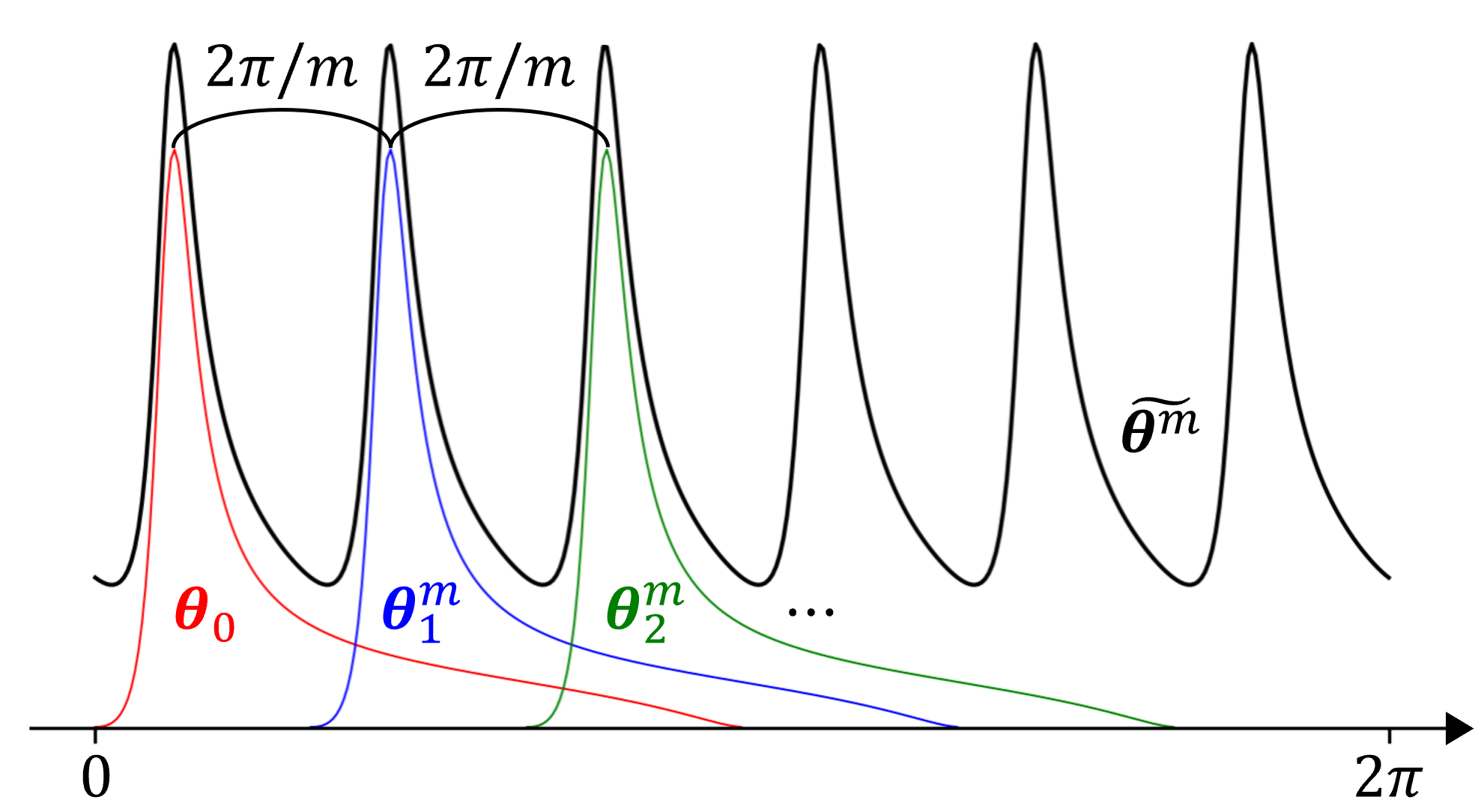}
    \caption{The example of distribution for a dataset $\tilde{\bm{\theta}}^m$ and $\bm{\theta}^m_k$.}
    \label{fig:periodic2}
\end{figure}

Let $\bm{\theta}_0=\{\theta_j\,|\,j=1,2,\dots,N\}$ denote the dataset of angles.
Now consider the dataset $\bm{\theta}^m_k=\{\theta_j+2\pi{k}/m\,|\,j=1,2,\dots,N\}$ where each data in $\bm{\theta}_0$ is rotated by $2\pi{k}/m~(m\in\mathbb{N}, m\geq{2}, k\in\mathbb{N}, k\leq{m-1})$ as shown in Figure~\ref{fig:periodic2}.
Then, the following lemma holds
\begin{lem}\label{lem3}
The sum of the dataset $\tilde{\bm{\theta}}^m=\bigcup_{k=1}^{m}\bm{\theta}^m_k$ has a period of $2\pi/m$.
\end{lem}

\noindent\textbf{Proof.} See \ref{proof3}.
\\
\\
Thus, showing $\nu_p{(\bigcup_{k=1}^{m}\bm{\theta}^m_k})=1,\ \forall{p\leq{m-1}}$ gives the proof of Lemma \ref{lem1}.
For that, we show $\bar{C}_p{(\bigcup_{k=1}^{m}\bm{\theta}^m_k)}$, $\bar{S}_p{(\bigcup_{k=1}^{m}\bm{\theta}^m_k)}=0,\ \forall{p\leq{m-1}}$.
Note that let $\bar{C}_p(\bm{\theta})$ and $\bar{S}_p(\bm{\theta})$ denote the cosine and sine components of the vector whose direction is the one for the composite of $(\cos p\theta_j, \sin p\theta_j)^{\top}$ for $j=1,2,\dots, N$, and length is divided by the number of data $N$, respectively.
\begin{align*}
\bar{C}_p{\Big(\bigcup_{k=1}^{m}\bm{\theta}^m_k\Big)}&=\frac{1}{m}\sum^{m}_{k=1}\Big(\frac{1}{n}\sum^{N}_{j=1}\cos{p}\big(\theta_j+\frac{2\pi{k}}{m}\big)\Big)\\
&=\frac{1}{mn}\sum^{N}_{j=1}\Big(\sum^{m}_{k=1}\cos{p}\big(\theta_j+\frac{2\pi{k}}{m}\big)\Big)\\
&=\frac{1}{mn}\sum^{N}_{j=1}\Big(\sum^{m}_{k=1}\big(\cos{p\theta_j}\cos{\frac{2\pi{kp}}{m}}-\sin{p\theta_j}\sin{\frac{2\pi{kp}}{m}}\big)\Big)\\
&=\frac{1}{mn}\sum^{N}_{j=1}\Big(\cos{p\theta_j}\sum^{m}_{k=1}\cos{\frac{2\pi{kp}}{m}}-\sin{p\theta_j}\sum^{m}_{k=1}\sin{\frac{2\pi{kp}}{m}}\Big),
\end{align*}
where the following lemma holds
\begin{lem}\label{lem4}
\
\begin{align}\label{cosp}
    \sum^{m}_{k=1}\cos\frac{2\pi{kp}}{m}=0,\ \sum^{m}_{k=1}\sin\frac{2\pi{kp}}{m}=0,\ \forall{p\leq{m-1}}.
\end{align}

\end{lem}

\noindent\textbf{Proof.} See \ref{proof4}.
\\
\\
Then, Eq.~(\ref{cosp}) gives
\begin{align*}
    \bar{C}_p{\Big(\bigcup_{k=1}^{m}\bm{\theta}^m_k\Big)}&=0,\ \forall{p\leq{m-1}}.
\end{align*}
$\bar{S}_p{(\bigcup_{k=1}^{m}\bm{\theta}^m_k)}=0,\ \forall{p\leq{m-1}}$ is also shown by the same procedure.
Therefore, $\nu_p{(\bigcup_{k=1}^{m}\bm{\theta}^m_k)}=1,\ \forall{p\leq{m-1}}$, i.e., $\nu_p(\tilde{\bm{\theta}}^m)=1,\ \forall{p\leq{m-1}}$ is proved.
\qed

\subsection{Proof of Lemma \ref{lem2}}\label{proof2}
As with Lemma \ref{lem1}, we show $\bar{C}_m{(\bigcup_{k=1}^{m}\bm{\theta}^m_k)}$, $\bar{S}_m{(\bigcup_{k=1}^{m}\bm{\theta}^m_k)}$ is not limited to $0$.
$\bar{C}_m{(\bigcup_{k=1}^{m}\bm{\theta}^m_k)}$ can be written down as
\begin{align*}
\bar{C}_m{\Big(\bigcup_{k=1}^{m}\bm{\theta^m_k}\Big)}&=\frac{1}{m}\sum^{m}_{k=1}\Big(\frac{1}{n}\sum^{N}_{j=1}\cos{m}\big(\theta_j+\frac{2\pi{k}}{m}\big)\Big)\\
&=\frac{1}{m}\sum^{N}_{j=1}\Big(\frac{1}{n}\sum^{m}_{k=1}\cos\big(m\theta_j+2\pi{k}\big)\Big)\\
&=\frac{1}{m}\sum^{N}_{j=1}\Big(\frac{1}{n}\sum^{m}_{k=1}\cos{m\theta_j}\Big).
\end{align*}
Now, the following lemma holds.

\begin{lem}\label{lem5}
The following equation holds
\begin{align*}
 \nu_p(\bm{\theta}\cup\bm{\theta})=\nu_p(\bm{\theta}).
\end{align*}
\end{lem}

\noindent\textbf{Proof.} Since $\bar{C}_p(\bm{\theta})$ and $\bar{S}_p(\bm{\theta})$ is the value divided by the number of data, and the direction does not change when data is replicated, $\bar{C}_p(\bm{\theta}\cup\bm{\theta})$ and $\bar{S}_p(\bm{\theta}\cup\bm{\theta})$ are equal to $\bar{C}_p(\bm{\theta})$ and $\bar{S}_p(\bm{\theta})$, respectively.
Therefore, $\nu_p(\bm{\theta}\cup\bm{\theta})$ is still equal to $\nu_p(\bm{\theta})$.
\qed
\\
\\
By Lemma \ref{lem5}, 
\begin{align*}
\bar{C}_m{\Big(\bigcup_{k=1}^{m}\bm{\theta}^m_k\Big)}=\bar{C}_m(\bm{\theta}_0).
\end{align*}
$\bar{S}_m{(\bigcup_{k=1}^{m}\bm{\theta}^m_k)}=\bar{S}_m(\bm{\theta}_0)$ is also shown by the same procedure.
Therefore, $\nu_m{(\bigcup_{k=1}^{m}\bm{\theta}^m_k)}$ (i.e., $\nu_m{(\tilde{\bm{\theta}}^m)}$) is not limited to $1$, but equal to $\nu_m{(\bm{\theta}_0)}$.
In addition, $\nu_m{(\bm{\theta}_0)}=0$ for only one data by its definition. Thus, $\nu_m(\tilde{\bm{\theta}}^m)$ is equal to 0  when $m$ data are equally distributed with an interval of $2\pi/m$,
\qed

\subsection{Proof of Lemma \ref{lem3}}\label{proof3}
$\bigcup_{k=1}^{m}\bm{\theta}^m_k$ can be written down as
\begin{align*}
    \bigcup_{k=1}^{m}\bm{\theta}^m_k=\Big\{&\theta_1+\frac{2\pi}{m},\theta_2+\frac{2\pi}{m},\dots,\theta_N+\frac{2\pi}{m},\\
    &\theta_1+2\pi\frac{2}{m},\theta_2+2\pi\frac{2}{m},\dots,\theta_N+2\pi\frac{2}{m},\\
    &\dots\\
    &\theta_1+2\pi\frac{m-1}{m},\theta_2+2\pi\frac{m-1}{m},\dots,\theta_N+2\pi\frac{m-1}{m}\Big\}.
\end{align*}
$\bigcup_{k=1}^{m}\bm{\theta}^m_k$ can be regarded as consisting of $N$ sets of $m$ data for each $\theta_j$, rotated by $2\pi{k}/m$, i.e., each $\theta_j$ has a period of $2\pi/m$,
Therefore, the set of $\bigcup_{k=1}^{m}\bm{\theta}^m_k$ (i.e., $\tilde{\bm{\theta}}^m$) also has a period of $2\pi/m$.
\qed

\subsection{Proof of Lemma \ref{lem4}}\label{proof4}
In the complex plane, all points of $\{\exp(2\pi{ikp}/m)\,|\,k=1,2,\dots,m\}$ are located at the vertices of a regular m-sided polygon whose all vertices are on the unit circle.
Note that let $i$ be the imaginary unit.
We use this property to prove Lemma \ref{lem4} by showing $\sum^m_{k=1}\exp(2\pi{ikp}/m)=0$.
First, we divide the case into two based on the relationship between $p$ and $m$.

\noindent(i) $p$ and $m$ are coprime

The all remainders of $p, 2p, \dots, mp$ divided by $m$ are different, i.e., there is no pair of $k'$ and $k''~(k',k''\in\mathbb{N}, k',k''\leq{m}, k'\neq{k''})$ that satisfies $k'p\equiv{k''p}\pmod m$.
Because $k'p\equiv{k''p}\pmod m$ can be transformed $k'\equiv{k''}\pmod m$, which means $k'=k''$ since $k',k''\leq{m}$, and this contradicts $k'\neq{k''}$.
Hence, all points of $\{\exp(2\pi{ikp}/m)\,|\,k=1,2,\dots,m\}$ are different, and they and all vertices of the regular m-sided polygon are one-to-one correspondence.
Since the center of gravity of the regular m-sided polygon is the origin, $\sum^m_{k=1}\exp(2\pi{ikp}/m)=0$.

\noindent(ii) $p$ and $m$ are not coprime

Let $l$ be the greatest common divisor of $p$ and $m$, and let $p', m'$ be a pair of coprime natural numbers satisfying $p=p'l, m=m'l$.
Then, $\{\exp(2\pi{ikp}/m)\,|\,k=1,2,\dots,m\}$ is equal to $\{\exp(2\pi{ikp'}/m')\,|\,k=1,2,\dots,m\}$.
Its sum is $0$ up to $k\leq{m'}$ as well as (i), and the same operation is repeated $l$ times.
Therefore, $\sum^m_{k=1}\exp(2\pi{ikp}/m)=0$ is held.

By (i) and (ii), $\sum^m_{k=1}\exp(2\pi{ikp}/m)=0,\ \forall p\in\mathbb{N}, p\leq{m-1}$ is shown and it means $\sum^{m}_{k=1}\cos(2\pi{kp}/m)$, $\sum^{m}_{k=1}\sin(2\pi{kp}/m)=0,\ \forall{p\leq{m-1}}.$
\qed

\section{Triangular pedestrian FD model}\label{ap2}

The proposed model Eq.~(\ref{eq12}) assumes that the flow is constant in the congested state for any value of density.
However, it is generally assumed that the flow decreases with increasing density in the congested state in vehicular FD.
To represent this phenomenon, a model using Newell's triangular FD \cite{newell2002simplified}, which is widely used in vehicular FDs, is verified.

The model to be estimated is as the following.
\begin{align}
    J(\rho,\nu_1, \nu_2, r)=-\log \{\exp(-u\rho) + \exp\left(-[1/\tau(1-\gamma_1\nu_1)(1-\gamma_2\nu_2)(1-\gamma_\mathrm{wall}r)+w\rho]\right)\}, \label{eqwithw}
\end{align}
where $\tau$ is the reaction time and $w$ is the backward wave speed.
A negative value of backward wave speed $w$ can represent a decrease in the flow in congested state.

Table \ref{suiteiw} shows the parameter estimation results and Table \ref{R2w} shows the value of the coefficients of determination $R^2$ and adjusted coefficients of determination $\bar{R}^2$ for the training and test data.
According to the estimation results in Table \ref{suiteiw}, $w$ is not significant and its sign is positive.
In addition, the value of adjusted coefficients of determination $\bar{R}^2$ for the test data is smaller than that of the proposed model in Table \ref{R2}.
Therefore, the proposed pedestrian FD model incorporating the angular variances does not need to represent a decrease in flow in the congestion state.

\begin{table}[htb]
    \centering
    \caption{Estimation results of the triangular FD.}
    \label{suiteiw}
    \begin{tabular}{c|crc}
    \hline
                         & Coefficient   & $t$ value & $p$ value              \\ \hline
    $u$      & 3.570  & 12.220 & $<10^{-10}$       \\
    $\tau$ & 0.658  & 11.523 & $<10^{-10}$       \\
    $\gamma_1$  & 0.293 & 7.008 & $<10^{-10}$ \\
    $\gamma_2$  & 0.243 & 2.744 & $6.791\times 10^{-3}$ \\
    $\gamma_{\mathrm{wall}}$    & 0.510 & 4.312 & $2.893\times 10^{-5}$ \\
    $w$    & 0.025 & 1.432 & 0.154 \\ \hline
    \end{tabular}
\end{table}

\begin{table}[htb]
    \centering
    \caption{Coefficient of determination of the triangular FD.}
    \label{R2w}
    \begin{tabular}{c|cc}
    \hline
           & $R^2$ & $\bar{R}^2$ \\ \hline
    Train data  & 0.668 & 0.655       \\
    Test data & 0.711 & 0.695     \\ \hline
    \end{tabular}
\end{table}

\section*{CRediT author statement contribution statement}

{\bf Kota Nagasaki}: Conceptualization; Formal analysis; Investigation; Methodology; Validation; Visualization; Writing - review \& editing.
{\bf Keiichiro Fujiya}: Conceptualization; Data curation; Formal analysis; Investigation; Methodology; Validation; Visualization; Writing - original draft.
{\bf Toru Seo}: Conceptualization; Funding acquisition; Methodology; Project administration; Resources; Supervision; Visualization; Writing - review \& editing.

\section*{Declaration of competing interest}

The authors declare that there are no known competing interest.

\section*{Declaration of generative AI and AI-assisted technologies in the writing process}

During the preparation of this work the authors used GPT-4 for a grammatical proofreading purpose.
After using this tool/service, the authors reviewed and edited the content as needed and takes full responsibility for the content of the publication.

\section*{Acknowledgement}

This work is partially supported by a JSPS KAKENHI Grant-in-Aid for Scientific Research 20H02267 and Grant-in-Aid for Challenging Research Exploratory 24K21656.

\bibliographystyle{elsarticle-num} 
\bibliography{MyLibrary.bib}

\begin{thebibliography}{10}
\expandafter\ifx\csname url\endcsname\relax
  \def\url#1{\texttt{#1}}\fi
\expandafter\ifx\csname urlprefix\endcsname\relax\def\urlprefix{URL }\fi
\expandafter\ifx\csname href\endcsname\relax
  \def\href#1#2{#2} \def\path#1{#1}\fi

\bibitem{vanumu2017fundamental}
L.~D. Vanumu, K.~Ramachandra~Rao, G.~Tiwari, Fundamental diagrams of pedestrian
  flow characteristics: A review, European transport research review 9 (2017)
  1--13.
\newblock \href {http://dx.doi.org/https://doi.org/10.1007/s12544-017-0264-6}
  {\path{doi:https://doi.org/10.1007/s12544-017-0264-6}}.

\bibitem{may1990traffic}
A.~D. May, Traffic flow fundamentals, Prentice Hall, 1990.

\bibitem{feliciani2016empirical}
C.~Feliciani, K.~Nishinari, Empirical analysis of the lane formation process in
  bidirectional pedestrian flow, Physical Review E 94~(3) (2016) 032304.
\newblock \href {http://dx.doi.org/https://doi.org/10.1103/PhysRevE.94.032304}
  {\path{doi:https://doi.org/10.1103/PhysRevE.94.032304}}.

\bibitem{mardia2009directional}
K.~V. Mardia, P.~E. Jupp, Directional statistics, John Wiley \& Sons, 2009.
\newblock \href {http://dx.doi.org/https://doi.org/10.1002/9780470316979}
  {\path{doi:https://doi.org/10.1002/9780470316979}}.

\bibitem{johnson1978some}
R.~A. Johnson, T.~E. Wehrly, Some angular-linear distributions and related
  regression models, Journal of the American Statistical Association 73~(363)
  (1978) 602--606.
\newblock \href
  {http://dx.doi.org/https://doi.org/10.1080/01621459.1978.10480062}
  {\path{doi:https://doi.org/10.1080/01621459.1978.10480062}}.

\bibitem{schnute1992statistical}
J.~T. Schnute, K.~Groot, Statistical analysis of animal orientation data,
  Animal behaviour 43~(1) (1992) 15--33.
\newblock \href
  {http://dx.doi.org/https://doi.org/10.1016/S0003-3472(05)80068-5}
  {\path{doi:https://doi.org/10.1016/S0003-3472(05)80068-5}}.

\bibitem{boeing2019urban}
G.~Boeing, Urban spatial order: Street network orientation, configuration, and
  entropy, Applied Network Science 4~(1) (2019) 1--19.
\newblock \href {http://dx.doi.org/https://doi.org/10.1007/s41109-019-0189-1}
  {\path{doi:https://doi.org/10.1007/s41109-019-0189-1}}.

\bibitem{nagasaki2023understanding}
K.~Nagasaki, T.~Seo, Understanding impact of angle in urban transportation,
  arXiv preprint arXiv:2310.16470\href
  {http://dx.doi.org/https://doi.org/10.48550/arXiv.2310.16470}
  {\path{doi:https://doi.org/10.48550/arXiv.2310.16470}}.

\bibitem{nagasakiApplicationRoseDiagram2019}
K.~Nagasaki, W.~Nakanishi, Y.~Asakura, Application of the rose diagram to road
  network analysis, in: 24th International Conference of Hong Kong Society for
  Transportation Studies, Hong Kong, 2019, pp. 169--176.

\bibitem{fruin1970designing}
J.~J. Fruin, Designing for pedestrians a level of service concept, Polytechnic
  University, 1970.
\newblock \href
  {http://dx.doi.org/https://doi.org/10.1016/B978-0-12-237502-8.50018-7}
  {\path{doi:https://doi.org/10.1016/B978-0-12-237502-8.50018-7}}.

\bibitem{virkler1994pedestrian}
M.~R. Virkler, S.~Elayadath, Pedestrian speed-flow-density relationships,
  Transportation research record 1438 (1994) 51--58.
\newblock \href {http://dx.doi.org/https://doi.org/10.3141/1438-08}
  {\path{doi:https://doi.org/10.3141/1438-08}}.

\bibitem{seyfried2005fundamental}
A.~Seyfried, B.~Steffen, W.~Klingsch, M.~Boltes, The fundamental diagram of
  pedestrian movement revisited, Journal of Statistical Mechanics: Theory and
  Experiment 2005~(10) (2005) P10002.
\newblock \href
  {http://dx.doi.org/https://doi.org/10.1088/1742-5468/2005/10/P10002}
  {\path{doi:https://doi.org/10.1088/1742-5468/2005/10/P10002}}.

\bibitem{zhang2013empirical}
J.~Zhang, A.~Seyfried, Empirical characteristics of different types of
  pedestrian streams, Procedia engineering 62 (2013) 655--662.
\newblock \href
  {http://dx.doi.org/https://doi.org/10.1016/j.proeng.2013.08.111}
  {\path{doi:https://doi.org/10.1016/j.proeng.2013.08.111}}.

\bibitem{lam2002study}
W.~H. Lam, J.~Y. Lee, C.~Cheung, A study of the bi-directional pedestrian flow
  characteristics at hong kong signalized crosswalk facilities, Transportation
  29 (2002) 169--192.
\newblock \href {http://dx.doi.org/https://doi.org/10.1023/A:1014226416702}
  {\path{doi:https://doi.org/10.1023/A:1014226416702}}.

\bibitem{jin2019observational}
C.-J. Jin, R.~Jiang, S.~Wong, S.~Xie, D.~Li, N.~Guo, W.~Wang, Observational
  characteristics of pedestrian flows under high-density conditions based on
  controlled experiments, Transportation research part C: emerging technologies
  109 (2019) 137--154.
\newblock \href {http://dx.doi.org/https://doi.org/10.1016/j.trc.2019.10.013}
  {\path{doi:https://doi.org/10.1016/j.trc.2019.10.013}}.

\bibitem{jia2021pedestrian}
X.~Jia, H.~Murakami, C.~Feliciani, D.~Yanagisawa, K.~Nishinari, Pedestrian lane
  formation and its influence on egress efficiency in the presence of an
  obstacle, Safety science 144 (2021) 105455.
\newblock \href {http://dx.doi.org/https://doi.org/10.1016/j.ssci.2021.105455}
  {\path{doi:https://doi.org/10.1016/j.ssci.2021.105455}}.

\bibitem{lee2016modeling}
J.~Lee, T.~Kim, J.-H. Chung, J.~Kim, Modeling lane formation in pedestrian
  counter flow and its effect on capacity, KSCE Journal of Civil Engineering 20
  (2016) 1099--1108.
\newblock \href {http://dx.doi.org/https://doi.org/10.1007/s12205-016-0741-9}
  {\path{doi:https://doi.org/10.1007/s12205-016-0741-9}}.

\bibitem{zhang2014comparison}
J.~Zhang, A.~Seyfried, Comparison of intersecting pedestrian flows based on
  experiments, Physica A: Statistical Mechanics and its Applications 405 (2014)
  316--325.
\newblock \href {http://dx.doi.org/https://doi.org/10.1016/j.physa.2014.03.004}
  {\path{doi:https://doi.org/10.1016/j.physa.2014.03.004}}.

\bibitem{cao2017fundamental}
S.~Cao, A.~Seyfried, J.~Zhang, S.~Holl, W.~Song, Fundamental diagrams for
  multidirectional pedestrian flows, Journal of Statistical Mechanics: Theory
  and Experiment 2017~(3) (2017) 033404.
\newblock \href {http://dx.doi.org/https://doi.org/10.1088/1742-5468/aa620d}
  {\path{doi:https://doi.org/10.1088/1742-5468/aa620d}}.

\bibitem{mihoPerformanceEvaluationPedestrian2015}
M.~Iryo, A.~Nagashima, Performance evaluation of pedestrian crossing flow,
  Seisan-kenkyu 67~(4) (2015) 369--373.
\newblock \href
  {http://dx.doi.org/https://doi.org/10.11188/seisankenkyu.67.369}
  {\path{doi:https://doi.org/10.11188/seisankenkyu.67.369}}.

\bibitem{helbing1995social}
D.~Helbing, P.~Molnar, Social force model for pedestrian dynamics, Physical
  review E 51~(5) (1995) 4282.
\newblock \href {http://dx.doi.org/https://doi.org/10.1103/PhysRevE.51.4282}
  {\path{doi:https://doi.org/10.1103/PhysRevE.51.4282}}.

\bibitem{asano2009pedestrian}
M.~Asano, T.~Iryo, M.~Kuwahara, A pedestrian model considering anticipatory
  behaviour for capacity evaluation, in: Transportation and Traffic Theory
  2009: Golden Jubilee, Springer, 2009, pp. 559--581.
\newblock \href
  {http://dx.doi.org/https://doi.org/10.1007/978-1-4419-0820-9$\_$28}
  {\path{doi:https://doi.org/10.1007/978-1-4419-0820-9$\_$28}}.

\bibitem{moussaid2009experimental}
M.~Moussa{\"\i}d, D.~Helbing, S.~Garnier, A.~Johansson, M.~Combe, G.~Theraulaz,
  Experimental study of the behavioural mechanisms underlying self-organization
  in human crowds, Proceedings of the Royal Society B: Biological Sciences
  276~(1668) (2009) 2755--2762.
\newblock \href {http://dx.doi.org/https://doi.org/10.1098/rspb.2009.0405}
  {\path{doi:https://doi.org/10.1098/rspb.2009.0405}}.

\bibitem{nakanishi2015preliminary}
W.~Nakanishi, T.~Fuse, A preliminary study on analysis of pedestrian traffic
  line using angular data, JSTE Journal of Traffic Engineering 1~(4) (2015)
  A$\_$18--A$\_$23.
\newblock \href
  {http://dx.doi.org/https://doi.org/10.14954/jste.1.4$\_$A$\_$18}
  {\path{doi:https://doi.org/10.14954/jste.1.4$\_$A$\_$18}}.

\bibitem{flotterod2015bidirectional}
G.~Fl{\"o}tter{\"o}d, G.~L{\"a}mmel, Bidirectional pedestrian fundamental
  diagram, Transportation research part B: methodological 71 (2015) 194--212.
\newblock \href {http://dx.doi.org/https://doi.org/10.1016/j.trb.2014.11.001}
  {\path{doi:https://doi.org/10.1016/j.trb.2014.11.001}}.

\bibitem{feliciani2018universal}
C.~Feliciani, H.~Murakami, K.~Nishinari, A universal function for capacity of
  bidirectional pedestrian streams: Filling the gaps in the literature, PloS
  one 13~(12) (2018) e0208496.
\newblock \href
  {http://dx.doi.org/https://doi.org/10.1371/journal.pone.0208496}
  {\path{doi:https://doi.org/10.1371/journal.pone.0208496}}.

\bibitem{Moustaid2021pedestrian}
E.~Moustaid, G.~Fl{\"o}tter{\"o}d, Macroscopic model of multidirectional
  pedestrian network flows, Transportation Research Part B: Methodological 145
  (2021) 1--23.
\newblock \href {http://dx.doi.org/https://doi.org/10.1016/j.trb.2020.12.004}
  {\path{doi:https://doi.org/10.1016/j.trb.2020.12.004}}.

\bibitem{flotterod2011node}
G.~Fl{\"o}tter{\"o}d, J.~Rohde, Operational macroscopic modeling of complex
  urban road intersections, Transportation Research Part B: Methodological
  45~(6) (2011) 903--922.
\newblock \href {http://dx.doi.org/https://doi.org/10.1016/j.trb.2011.04.001}
  {\path{doi:https://doi.org/10.1016/j.trb.2011.04.001}}.

\bibitem{saberi2014exploring}
M.~Saberi, H.~S. Mahmassani, Exploring areawide dynamics of pedestrian crowds:
  three-dimensional approach, Transportation research record 2421~(1) (2014)
  31--40.
\newblock \href {http://dx.doi.org/https://doi.org/10.3141/2421-04}
  {\path{doi:https://doi.org/10.3141/2421-04}}.

\bibitem{edie}
L.~Edie, Discussion of traffic stream measurements and definitions, in: 2nd
  International Symposium on the Theory of Traffic Flow, {London}, 1963, pp.
  139--154.

\bibitem{wang2019fundamental}
P.~Wang, S.~Cao, M.~Yao, Fundamental diagrams for pedestrian traffic flow in
  controlled experiments, Physica A: Statistical Mechanics and its Applications
  525 (2019) 266--277.
\newblock \href {http://dx.doi.org/https://doi.org/10.1016/j.physa.2019.03.057}
  {\path{doi:https://doi.org/10.1016/j.physa.2019.03.057}}.

\bibitem{nielsen2016guaranteed}
F.~Nielsen, K.~Sun, Guaranteed bounds on information-theoretic measures of
  univariate mixtures using piecewise log-sum-exp inequalities, Entropy 18~(12)
  (2016) 442.
\newblock \href {http://dx.doi.org/https://doi.org/10.3390/e18120442}
  {\path{doi:https://doi.org/10.3390/e18120442}}.

\bibitem{newell2002simplified}
G.~F. Newell, A simplified car-following theory: a lower order model,
  Transportation Research Part B: Methodological 36~(3) (2002) 195--205.
\newblock \href
  {http://dx.doi.org/https://doi.org/10.1016/S0191-2615(00)00044-8}
  {\path{doi:https://doi.org/10.1016/S0191-2615(00)00044-8}}.

\end{thebibliography}

\end{document}